\documentclass[traditabstract]{aa} 
\usepackage{graphicx}
\usepackage{txfonts}

\usepackage{natbib} 
\bibpunct{(}{)}{;}{a}{}{,} 
\usepackage{multirow}

\def \th {\thinspace}

\def \ergsc{\hbox{erg s$^{-1}$ cm$^{-2}$}}

\def \ergs{\hbox{erg s$^{-1}$}}

\def \nh {\hbox{ $N{\rm _H}$ }}
\def \colc {cm$^{-2}$}

\def \fek {Fe K$\alpha$}
\def \chidof{$\frac{\chi ^{2}}{\rm d.o.f}$}

\usepackage{amstext}

\begin{document}

\title{Prospects for IXPE and
eXTP polarimetric archaeology of the reflection nebulae in the Galactic center}
\author{L. Di Gesu \inst{1}
 \and R. Ferrazzoli \inst{2}
 \and I. Donnarumma \inst{1}
 \and P. Soffitta \inst{2}
 \and E. Costa \inst{2}
 \and F. Muleri \inst{2}
\and  M. Pesce-Rollins \inst{3}
\and  F. Marin \inst{4}}
\institute
{Italian Space Agency (ASI), Via del Politecnico snc, 00133, Roma, Italy
\and
INAF/IAPS, via del Fosso del Cavaliere 100, 00133, Roma, Italy
\and
Istituto Nazionale di Fisica Nucleare, Sezione di Pisa, I-56127 Pisa, Italy
\and
Universit\'e de Strasbourg, CNRS, Observatoire astronomique de Strasbourg, UMR 7550, F-67000 Strasbourg, France}
\date{}
\abstract{
The X-ray polarization properties of the reflection nebulae in the Galactic center inform us
about the direction of the illuminating source (through the polarization angle) and
the cloud position along the line of sight (through the polarization degree). 
However, the detected polarization degree is expected to be lowered because the polarized emission of the clouds is mixed with the unpolarized diffuse emission 
that  permeates the Galactic center region. In a real observation, also the morphological smearing
of the source due to the point spread function and the unpolarized instrumental background 
contribute in diluting the polarization degree. So far, these effects have never been included
in the estimation of the dilution.\\
We evaluate the detectability of the X-ray polarization
predicted for the MC2, Bridge-B2, G0.11-0.11, Sgr B2, Sgr C1,  Sgr C2, and Sgr C3 molecular clouds
with modern X-ray imaging polarimeters  such as the Imaging X-ray Polarimetry Explorer (IXPE), which is expected 
to launch in 2021, and the Enhanced X-ray Timing and Polarimetry mission (eXTP), whose launch is scheduled for 2027. 
We perform realistic simulations of X-ray polarimetric observations considering (with the aid
of Chandra maps and spectra) the spatial, spectral, and polarization 
properties of all the diffuse emission  and background components in each region of interest. \\
We find that in the 4.0$-$8.0 keV band,  where the emission of the molecular clouds outshines the other components,
the dilution of the polarization degree, including the contribution due
to the morphological smearing of the source, ranges between $\sim$19\% and $\sim$55\%. 
We conclude that for some distance values reported in the
literature, the diluted polarization degree
of G0.11-0.11, Sgr B2, Bridge-B2, Bridge-E,  Sgr C1, and
Sgr C3 may be detectable in a 2 Ms long IXPE observations.
With the same exposure time, and considering the whole
range of possible distances reported in the literature, the enhanced capabilities 
of eXTP may allow detecting the 4.0$-$8.0 keV of
all the targets considered here.}
\titlerunning{Prospects for
polarimetric archaeology of the reflection nebulae in the Galactic center}
\authorrunning{Di Gesu L. et al}
\maketitle
%
%
\begin{figure}
\includegraphics[width=0.5\textwidth]{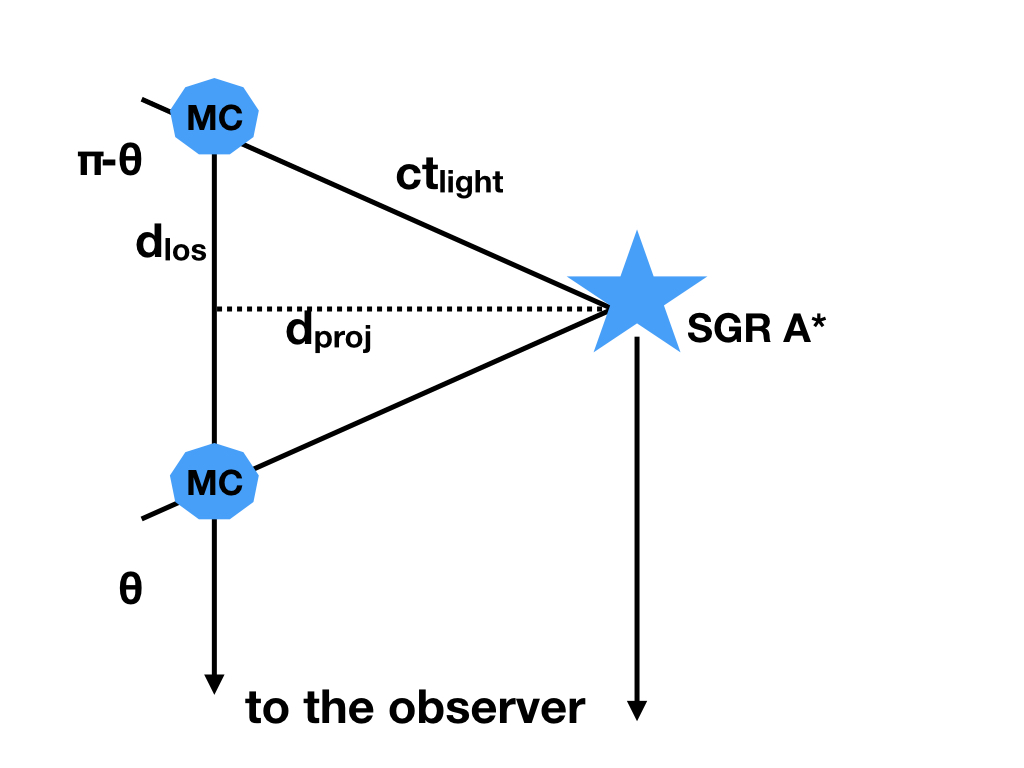}
\caption{Scattering geometry for
a molecular cloud located in front
or behind the Sgr A* plane. The
two positions result in
the same effective scattering
and polarization degree.
In this scheme, $d_{\rm proj}$ is the
cloud-Sgr A* distance projected in the plane
of the sky, $d_{\rm los}$ is the line-of-sight
displacement of the cloud with respect
to the Sgr A* plane, c is the speed of light,
$t_{\rm light}$ is the light travel time between
Sgr A* and the cloud, $\theta$ and $\pi-\theta$
are the two possible scattering
angles.}
 \label{geo.fig}
\end{figure}
\section{Introduction}
\label{intro}
%
%
The supermassive black hole (SMBH) Sgr A* that today lies in the center 
of our Galaxy, is a low-luminosity,
X-ray dim ($L_X \sim 2 \times 10^{33}$ \ergs, 
\citealt{baganoff2001}) galactic nucleus. 
Nonetheless,
some observed phenomena
in the Galactic center (GC) region
are explained by
past more luminous
phases of Sgr A*
\citep[see][for a review]{ponti2013}. 
For instance, 
the  huge gamma-ray bubble
that Fermi-LAT observed
10  kpc  above  and  below  the  GC
may be the remnant of an active phase
of Sgr A* a few  million  years  ago
\citep{su2010, zubovas2011}.  
Determining the history
of the activity of
our Galactic nucleus would
allow us to assess the duty cycle
of mass accretion of the SMBH
\citep[e.g.,][]{parkho2012} and thus provide unique
insight into the coevolution
of the SMBH and galaxies 
\citep{dimatteo2008}.\\
The quest
of reconstructing
the history of Sgr A*
has motivated the interest in characterizing
the central molecular zone
\citep[CMZ,][]{morris1996},
the $\sim$ 100 pc extended region around Sgr A*.
The CMZ hosts several molecular cloud complexes
(e.g., Sgr A,
Sgr B, and Sgr C) that are visible, for instance,
in the thermal far-infrared images
obtained with the
Herschel satellite \citep{molinari2011}.
Interestingly,
the physical conditions in the CMZ
inferred from infrared observations
(i.e., the geometrical size, column
density, and gas dynamics) are reminiscent
of an AGN torus \citep{ramos2017}.\\
The molecular gas in the CMZ
is also traced by 
X-ray reflection spectral features, 
such as a prominent \fek \, line 
and a reflection continuum
\citep{ponti2013}.
The lack of X-ray bright sources nearby
led \citet{sunyaev1993} 
to suggest 
that the observed emission is the echo of an outburst of Sgr A* that occurred a
$ \text{}$few hundred years ago and reached a peak luminosity of $10^{39-40}$ \ergs.
According to this scenario, the reflected
radiation is still visible
because of the delay induced
by the light travel time between Sgr A*
and the clouds in the CMZ. \\
We sketch two
possible scattering
geometries of an individual
cloud  located in front or behind
the Sgr A* plane
in Fig. \ref{geo.fig}.
Hereafter,  we indicate
with $d_{\rm proj}$
the distance between the
cloud and Sgr A* projected
on the plane of the sky, with
$d_{\rm los}$ the  line-of-sight  
displacement  of the cloud
with respect  to the Sgr  A*  plane,
and
$\theta$ the scattering
angle. In addition,
c is the speed of light
and $t_{\rm light}$ 
the light travel time between
Sgr A* and the cloud. 
The two positions depicted
in Fig. \ref{geo.fig} result
in the same effective scattering.\\
The hypothesis of a previous Sgr A*  outburst is appealing
because it implies that
the X-ray variability of the
CMZ is a fossil memory 
of how our Galactic nucleus
acted $\text{a }$few hundred
years ago \citep{muno2007}. 
Over the years, great effort
has been devoted to reconstructing
the past light curve of Sgr A*
using X-ray spectral, timing,
and imaging techniques
\citep{koyama1996, murakami2001}.
A single-outburst scenario \citep{ponti2010}, 
a two-burst scenario \citep{clavel2013},
and a number of short-lived events \citep{terrier2018}
have been suggested to explain
the data. These flares
may be superposed to
a long-term high state
of Sgr A* \citep{ryu2013}.\\
The main source
of uncertainty in
these studies
is that $d_{\rm los}$ 
is poorly constrained,
which makes it difficult
to infer the time delay 
$t_{\rm light}$
and the number
of illuminating events.
So far, two methods
have been used to
overcome this problem.
Some works searched
for correlating variations in
multiple regions throughout the
CMZ, which
provides indications
on the number
and nature
of the illuminating events
\citep{clavel2013, churazov2017, terrier2018}.
Conversely,
other authors
have attempted to
derive the line-of-sight positions of individual clumps
from a detailed modeling
of the iron line
and of the reflection continuum
\citep{capelli2012, walls2016, chuard2018}.
These reflection models
assume a geometry
in which the illuminating
source is in Sgr A*,
which is still debated.
Although disfavored
as an explanation
for the steady part of the emission, alternative sources
of illumination, such as cosmic rays
from a local source penetrating
the clouds  \citep{yusef-zadeh2013,dogiel2014},
are not conclusively
ruled out by current data
\citep{mori2015,
zhang2015}.\\
An independent way to address these ambiguities
is provided by X-ray polarimetry.
The reflected emission from a compact illuminating, 
source
is linearly polarized by scattering in the absence
of depolarizing agent. 
The expected polarization
angle is normal to the scattering
plane and  therefore carries clean information of the direction
of the illuminating source.
The expected
polarization degree P
depends on the
scattering angle $\theta$
\citep{mcmaster1961}
by
\begin{equation}
P= \frac{1-\cos^2 \theta}{1+\cos^2 \theta}
\label{eqp}
.\end{equation}
Thus, a measurement
of the polarization
degree  of a molecular
cloud allows us to determine
$d_{\rm los}$ because according
to the geometry of
Fig. \ref{geo.fig},
\begin{equation}
d_{\rm los}=d_{\rm proj}\cot\theta
\label{eqd}
.\end{equation}
The remaining
ambiguity of whether
$d_{\rm los}$ is positive
or negative
can be broken, for instance,
using spectral
information
(i.e., the dependence
of the equivalent
width of the iron line
on the scattering angle
and iron abundances,
see \citealt{churazov2017}).
An X-ray polarization study
of the molecular clouds
in the GC has the potential
of addressing the critical
uncertainties 
that still
hamper a full understanding
of the origin of
the reflection of the
nebulae in the GC
\citep{churazov2002, 
marin2014, marin2015,
churazov2017}.\\
A physical limit of this experiment 
is the fact that the molecular clouds 
are embedded in the diffuse
unpolarized emission of
the GC region
\citep{koyama1989, sidoli1999}.
In addition to the X-ray reflection 
from the molecular clouds, the $2 - 8$ keV emission 
in the GC region comprises the contribution 
of two diffuse emission components 
\citep[see][and references therein]{ponti2013}
that hereafter we call soft 
and hard plasma.  
The soft plasma
is traced by the \ion{Si}{xii},
\ion{Si}{xiii}, \ion{S}{xv}, and \ion{Ar}{xvii} lines, for example.
They are ascribed to a $\sim$1 keV,
collisionall -ionized
plasma
that pervades the GC
region and can be sustained by the supernova activity in the
region.
Conversely, hard plasma is traced
by a \ion{Fe}{xxv}-He$\alpha$ line emission
at $\sim6.7$ keV 
that morphologically
peaks in the central degree.
This component is often modeled 
as $\sim$6.5 keV thermal
plasma. At least a part of
it may be ascribed to unresolved
point sources such as
accreting white dwarf and coronally
active stars \citep{revnivtsev2007, yuasa2012}.
The remaining emission
might be associated with truly diffuse
hot gas, possibly originating
from supernova remnants.\\
Because of the complexity of the diffuse
emission in the GC region,
the synergy between polarimetric and imaging capabilities 
is a crucial asset for this study because it allows us
to resolve the faint molecular clouds 
from the diffuse emission in the background.
The NASA/ASI Imaging X-ray Polarimetry Explorer 
(IXPE, \citealt{weisskopf2016}) that will be launched in 2021 
is the first mission that is entirely dedicated to 
X-ray polarimetry through imaging-capable detectors
(i.e., gas pixel detector, GPD, \citealt{gpd}) 
in the $2 - 8$ keV band, 
and it will offer the first opportunity 
to investigate the X-ray polarization
of the GC region. The 
Enhanced X-ray 
Timing and Polarimetry 
mission \citep[eXTP,][]{extp}, 
which is planned 
to launch in 2027,
will also carry
a GPD polarimeter.
The effective area
of eXTP is expected
to be larger by
a factor $\sim$4 than that
of IXPE, and will
therefore allow enlarging
the pool of suitable
targets. \\
We evaluate the detectability of the X-ray polarization 
predicted for the molecular clouds 
in the GC region
by \citet{marin2015}.
We simulate IXPE observations
of individual
candidate targets.
With the aid
of Chandra maps and spectra,
we consider
(when possible)
the polarization, spectral,
and spatial properties of all
the emission components
(i.e., the cloud, soft plasma,
and hard plasma)
in each target field.
Chandra images
are most suitable
for this work because the
spatial resolution of Chandra
is infinite from the IXPE
point of view.
In addition, we include  a realistic model
for the  instrumental
background and for the
cosmic X-ray background
(CXB). In this way,
we are able to 
quantify how much
the polarization
degree of the molecular clouds
is diluted in the unpolarized
environmental radiation.
In the ideal case of
a detector with an infinite
spatial resolution and
zero background, the dilution
factor is just the ratio
between the reflection flux
and the total flux \citep[as in, e.g.,][]{marin2015}.
In a real observation, 
the morphological smearing
due to a finite point
spread function (PSF)
and the unpolarized
background 
contribute to
increasing the dilution.
Our simulation strategy
allows us to quantify
this additional dilution as well.\\
Throughout this paper, we quantify the detectability 
of the polarization by computing the minimum detectable polarization (MDP). 
The MDP \citep{weisskopf2010} is the fundamental
quantity for the statistical significance 
of an X-ray polarization measurement and is defined as
\begin{equation}
\rm MDP_{99\%} = \frac{4.29}{\mu R} \sqrt{\frac{R + B}{T}}
\label{mdp}
,\end{equation}
where $R$ is the detected source rate (in counts/s), 
$B$ is the background rate, $T$ is the observation time (in seconds), 
and $\mu$ is the adimensional modulation factor of the detector.
The $\rm MDP_{99\%}$ is not the uncertainty of the polarization measurement, 
but rather the degree of polarization that can be determined
with a 99\% probability against the null
hypothesis.\\
%
The paper is organized as follows.
In section \ref{ch} we describe
the selection and preparation
of the Chandra data,
and in section \ref{stra} 
we present our simulation procedure.
Finally, in section \ref{disc} 
we discuss the results,
and in Sect. \ref{conc} 
we summarize our conclusions.
%
%
%
%
\begin{figure*}
 \includegraphics[width=1.0\textwidth]{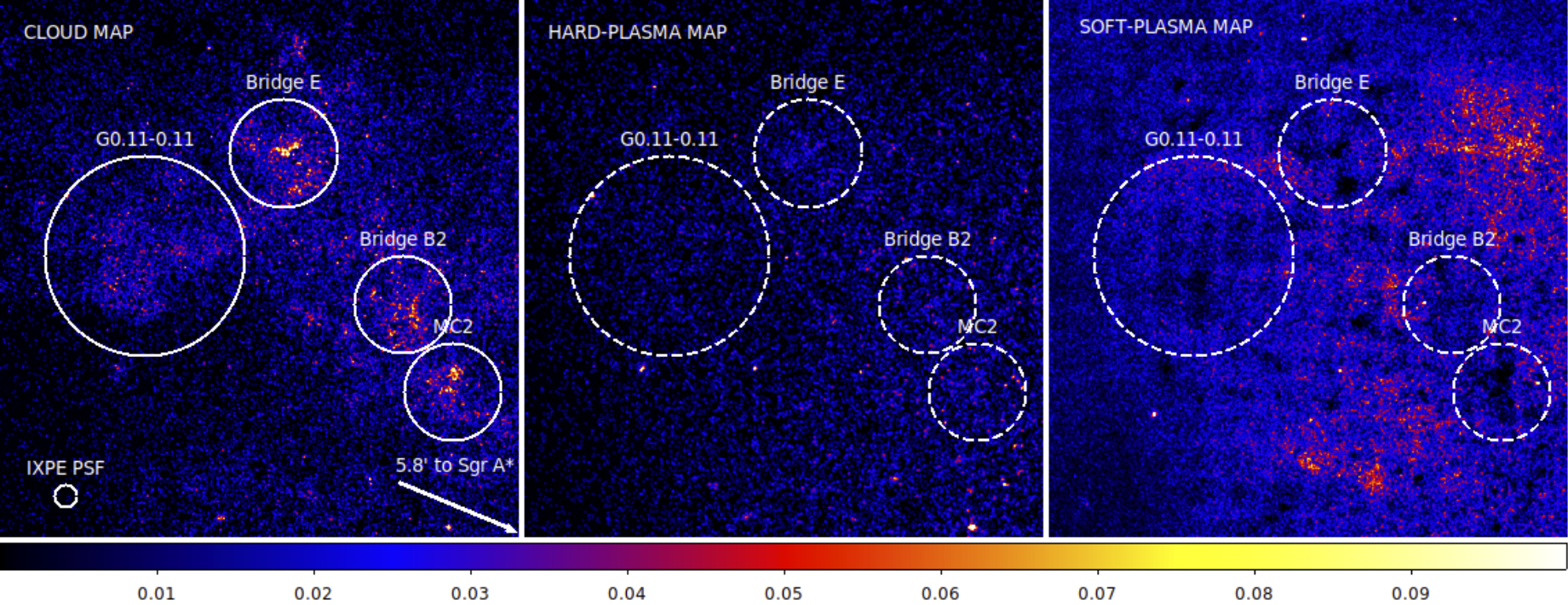}
  \caption{From left to right: Background-
  and continuum-subtracted Chandra maps of the cloud, hard plasma,
  and soft plasma component in the Sgr A region. Images are smoothed
  using a 3-pixel Gaussian kernel. The color bar displayed
  at the bottom has adimensional units because the images
  are normalized to the maximum value.
  The regions comprising the targets selected
  for IXPE simulations (i.e., MC2, Bridge-B2, Bridge-E, and
  G0.11-0.11) are shown.
  In the first panel, a circle with the size
  of the IXPE PSF is shown for comparison. The direction
  of Sgr A* is indicated with an arrow.}
  \label{sga_maps.fig}
\end{figure*}
%
%
%
%
\section{Chandra data preparation}
\label{ch}

\subsection{Chandra data selection}
\label{datasel}
We consider as candidate
targets for an X-ray polarimetry
observation
the molecular clouds for which 
\citet {marin2015} computed
the polarization properties expected
in a theoretical scenario where 
the source of illumination
is a past unpolarized outburst of Sgr A$^{*}$.
The molecular clouds
MC1, MC2, Bridge-D,
Bridge-E, Bridge-B2
and G0.11-0.11
belong to  the
Sgr A complex.
The Sgr B complex
comprises two
substructures
named Sgr B1 and
Sgr B2.
Conversely,
the clouds Sgr C1, Sgr C2,
and Sgr C3 are substructures
of the Sgr C complex.\\
The morphology of the molecular
clouds is known from extensive
Chandra and XMM-Newton observational
campaigns that were carried
out in the past 20 years.  
The extension of the clouds
 \citep[see, e.g.,][]{terrier2018}
is typically larger than the nominal
PSF of IXPE (which has
a radius of $\sim$10\arcsec).
Furthermore,
the diffuse plasma in which the clouds
are embedded has an inhomogeneous
morphology.
In some of 
the simulations that follow,
we therefore use Chandra maps 
to define the extended
spatial morphology of
the cloud and the soft
and hard plasma 
component. Moreover,
Chandra spectra
are used to input the spectral
shape of each emission
component.\\
As a first step in
the preparation of
the IXPE simulations,
we retrieved
from the public archive
the Chandra observations
of the Sgr A, Sgr B, and Sgr C
complexes.
We selected
in the archive all the Chandra ACIS-I
observations that were taken since 1999
without any gratings in place.\\
For the Sgr A field,
the total Chandra
exposure time is
$\sim$ 2.4 Ms. 
Owing to the superior quality
of this dataset,
we were able to compute
Chandra images of
the cloud and the soft
and hard plasma
in this region
(Sect. \ref{chmap}).
The Chandra field of Sgr B
comprises only Sgr B2, and no Chandra observations
include Sgr B1, which
is therefore excluded from
our analysis.
The Chandra
field of Sgr C comprises Sgr C1
and Sgr C2, and 
Sgr C3 is included
in Chandra
Obs-ID 7040.
For the Sgr B and Sgr C
region the total
exposure time of the
data available
in the Chandra archive
is insufficient
to produce sensible
maps of the emission
components separately.
For these clouds
we therefore use the most recent available
Chandra observation for the spectral
analysis
(Sect. \ref{chspec}).
We list
in Table \ref{obs.tab}
all the Chandra 
observations that we used.
%
%
%
%
%
\begin{table*}
\caption{Input data for IXPE simulations of molecular clouds.}     
\label{input.tab}      
\centering                    
\begin{tabular}{lccc}        
\hline\hline                 
Region\tablefootmark{a}
& Identification \tablefootmark{b}
& $d_{\rm los}$ \tablefootmark{c} 
& Polarization properties \tablefootmark{d}\\
\hline
%
%
Center,  \quad \,\,\, \,\,\ \,\,\, \,\,\, \,\,\ \,\,\, \, radius, \quad $ d_{\rm proj}$
&
& 
& P \quad $\phi$\\
%
%
(hh:mm:ss.s, dd:mm:ss.s, \quad \arcsec \quad pc)
&
& (pc)
& (\%) \quad (\degr) \\
\hline
%
%
17:46:00.6,-28:56:49.2, \quad 49 \quad -14
& MC2
& -17
&$25.8\%$ \quad $73.8\degr$\\
%
%
%
17:46:05.5,-28:55:40.8 \quad 44 \quad -18
& Bridge B2
& -60
&$15.8\%$ \quad $77.8\degr$\\
%
%
%
17:46:12.1,-28:53:20.3 \quad 49 \quad -25
& Bridge E
& -60
&$12.7\%$ \quad $67.9\degr$\\
%
%
%
%
17:46:21.6,-28:54:52.1 \quad 90 \quad  -27
& G0.11-0.11
& -17
&$55.8\%$ \quad $61.6\degr$\\
\hline
%
%
%
%
%
%
17:47:30.60,-28:26:36.6 \quad 121 \quad -100
&Sgr B2
& -17
& $65.0\%$ \quad $88.3\degr$ \\
\hline
%
 17:44:30.63, -29:27:22.6 \quad 100 \quad 71
& Sgr C1
& -74
& $31.1\%$ \quad $94.6\degr$\\
%
%
17:44:54.93, -29:28:30.4 \quad 115 \quad 66
& Sgr C2
& 58
& $34.9\%$ \quad $99.1\degr$\\
%
%
%
17:45:12.19, -29:22:22.0 \quad 146 \quad 50
& Sgr C3
& -53
& $32.9\%$ \quad $106.4\degr$\\
\hline
\end{tabular}
\tablefoot{
\tablefoottext{a}{Data of the regions for the spectral analysis and IXPE simulations. Positive and negative projected distances mean east and west of the GC.}
\tablefoottext{b}{Cross identification with the target names used in \citet{marin2015}.}
\tablefoottext{c}{Distance along the line of sight assumed in \citet{marin2015}. See references therein. Positive and negative means behind and in front
of the Galactic plane.}
\tablefoottext{d}{Polarization properties from the model of \citet{marin2015}.}}
\end{table*}

\subsection{Chandra maps of the Sgr A field}
\label{chmap}
We processed the Chandra data
using the CIAO software \citep{fruscione2006},
version 4.11, in combination
with version 4.8.2
of the Chandra calibration
database (CALDB).
For each observation,
we ran the chandra\_repro
routine to create the clean
level 2 event file.
Hence, for 
the Sgr A region,
we created 
background
and continuum-subtracted 
counts maps of the soft and hard plasma, and the clouds.
For all the images, we kept the native ACIS 
pixel size (i.e., $\sim$0.5\arcsec).\\
We proceeded as follows. 
For each observation,
we created the background event-file
using the blank-sky
event files that are provided 
in the Chandra CALDB. For this,
we used the blanksky CIAO routine,
which customizes a blanksky 
background file for the 
input event file, finding
the instrument-specific 
background files in the CALDB
and combining and reprojecting 
them to match the input 
coordinates.\\
For each observation
we ran the blanksky-image script
to create background-subtracted
Chandra count maps of each emission
component.
For the soft plasma, we created
a map in the 2.35$-$3.22 keV energy
band, which comprises the
\ion{S}{xv} and
\ion{Ar}{xvii}
emission lines.
For the hard plasma,
we created a map centered
on the \ion{Fe}{xxv}-He$\alpha$
line (6.62$-$6.78 keV). The morphology
of the molecular gas is
given by a Chandra map centered
on the Fe-K$\alpha$ line (6.32$-$6.48 keV).
Finally, for the continuum
we used the 4.0$- $6.32 keV
band \citep[e.g.,][]{clavel2013},
which is line free.\\
For each emission
component, we merged as a final step all
the images using the 
CIAO script
reproject\_image\_grid routine,
which reprojects all the
input images to a common coordinates
grid. Using the spectra
of the four targets that
we selected for the simulations
(see the following
Section \ref{chtarget}
and 
Section \ref{chspec}),
we found
that a model
including an absorbed
power law and a 1 keV
brehmsstrahlung
component adequately
interpolate the continuum spectral shape
that underlies the \ion{S}{xv},
\ion{Ar}{xvii},
\ion{Fe}{xxv}-He$\alpha,$
and Fe-K$\alpha$ line.
By averaging the results
of this continuum model
for the four targets of interest,
we derived the scaling factors
(0.38, 0.10, and 0.09 for
the soft and hard plasma,
and the cloud band, respectively)
that we used to rescale
the continuum images
in the band of
each emission
component. Thus,
these scaling factors
were optimized
for the regions
we used the simulations
that followed.
The final
images of each emission component
were obtained by subtracting
the rescaled continuum count-maps
from the signal count maps.
We normalized all the
maps by dividing
by the maximum value.
We display the final
background-
and continuum-subtracted
maps of the three
emission components
in Fig \ref{sga_maps.fig}. \\
\subsection{Target identification in the Sgr A field
and Chandra maps of individual targets}
\label{chtarget}
We searched for the targets analyzed
in \citet{marin2015}
in the background- and
continuum-subtracted
\fek \, map of the Sgr A field
(Fig. \ref{sga_maps.fig}, first
panel). We excluded
from our search and thus
from the IXPE simulations MC1 and the Bridge
D cloud because they are predicted
to be basically unpolarized.
We identified MC 2, Bridge B2,
Bridge E, and G0.11-0.11,
which are displayed
as circular regions
in  Fig. \ref{sga_maps.fig}. In Table 
\ref{input.tab} we list the
central coordinates,
the radius, and the
projected distance
from Sgr A* of each cloud.
The cloud sizes are the same
of \citet{marin2015}.
As a final step in the
preparation of the maps
for the simulations
of the MC2, Bridge B2, Bridge E,
and G0.11-0.11 clouds,
we created
for each emission component
smaller Chandra maps
cut in the region of interest
(i.e., the
region listed in Table \ref{input.tab}).
This is because we simulated
IXPE observations
of each target individually
and on axis. We note,
however, that the IXPE
field of view is 9\arcmin
          in  radius, and thus a single
IXPE pointing of the
Sgr A field will
catch more than
one target. A simulation
mapping the entire 
IXPE field of view
will be presented in a
future expansion of this work.
Here, we simulate each cloud
individually, with the aim
of collecting useful
information
in order to decide the best target
for a pointing.\\
We centered each map
on the brightest Fe K$\alpha$
patch.  Because the morphology
of the clouds varies with
time, these coordinates
are shifted with respect
to those used
in \citet{marin2015}. 
This does not affect
the expected polarization
degree because it
depends mainly
on the
galactic depth
(Eq. \ref{eqp}).
The expected
polarization angle
may be affected,
but changes are
expected to be
less than one degree
(F. Marin, 
private communication).\\
In the case
of Sgr B1, Sgr C1,
Sgr C2, and Sgr C3,
we were unable to create the Fe-K$\alpha$ map
to search for
the cloud positions. For these clouds
we therefore used
the same regions
as \citet{marin2015}
to extract the spectra
from the most recent
Chandra observations.
The regions used for
Sgr B2, Sgr C1, Sgr C2,
and Sgr C3 are also listed
in Table \ref{input.tab}.\\
Finally, we list
in Table \ref{input.tab}
all the other cloud data
that we input in the IXPE
simulations, that is,
the polarization degrees
and angle resulting from
the model of \citet{marin2015}
that were computed
assuming a position 
$d_{\rm los}$
along the line of sight
of the clouds. The
assumed distance
is the key parameter
determining the polarization
degree and hence the
IXPE detectability.
We explore the effect
of the assumed
distances for
our simulations
in Sect. \ref{disc}.\\
%
%
%

%
%
\begin{figure*}
\begin{minipage}[c]{0.9\textwidth}
 \includegraphics[width=0.5\textwidth]{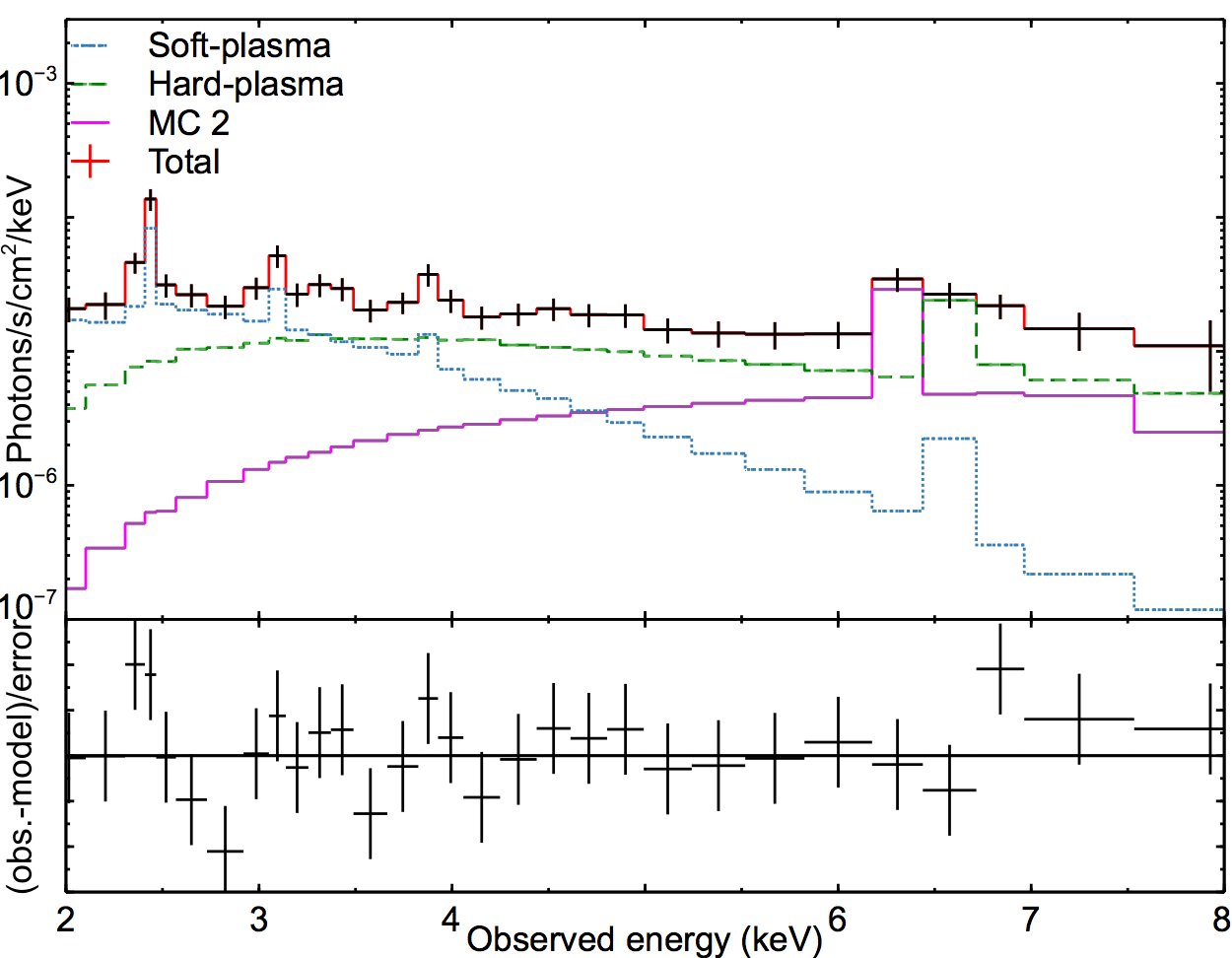}
 \includegraphics[width=0.5\textwidth]{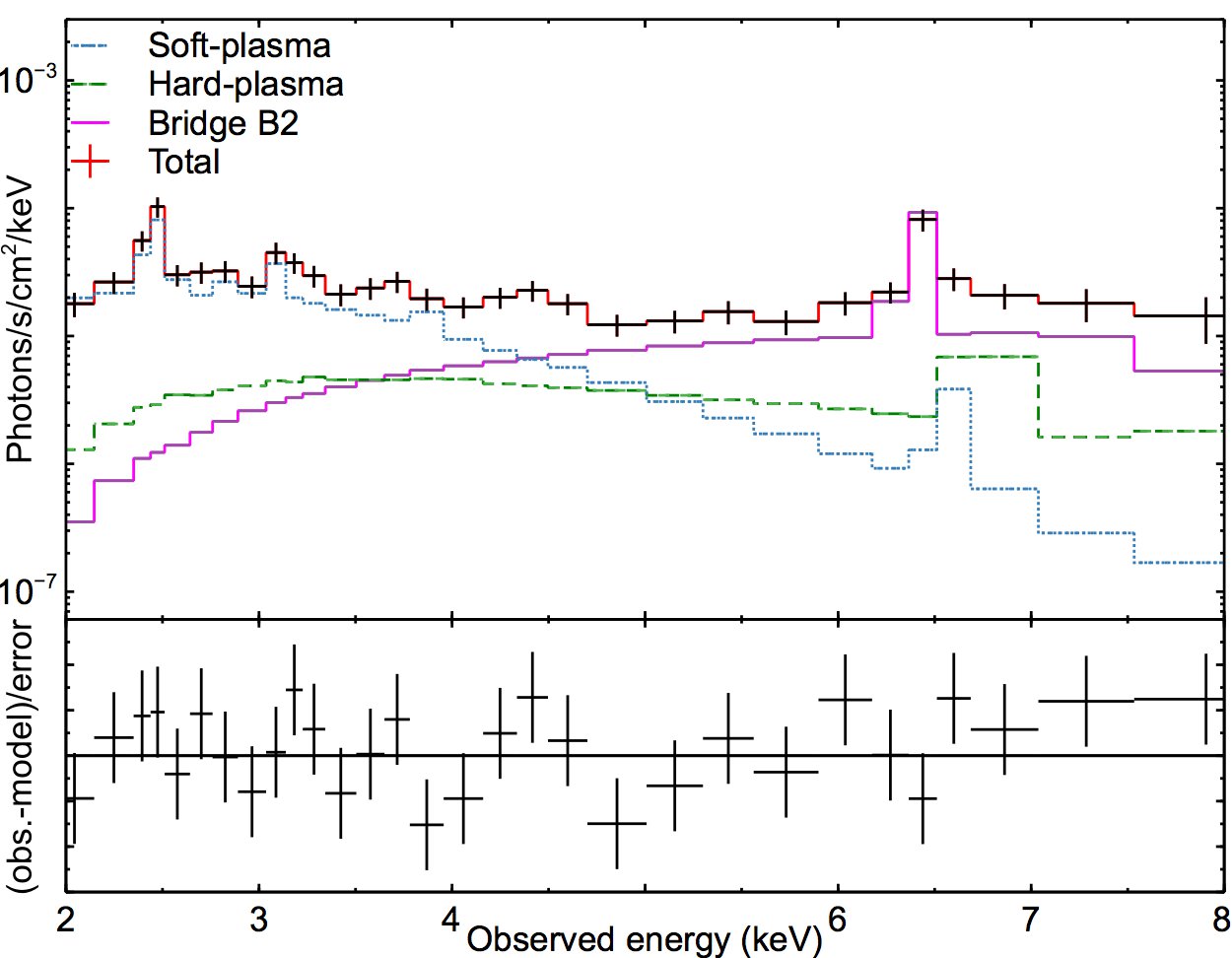}
 \includegraphics[width=0.5\textwidth]{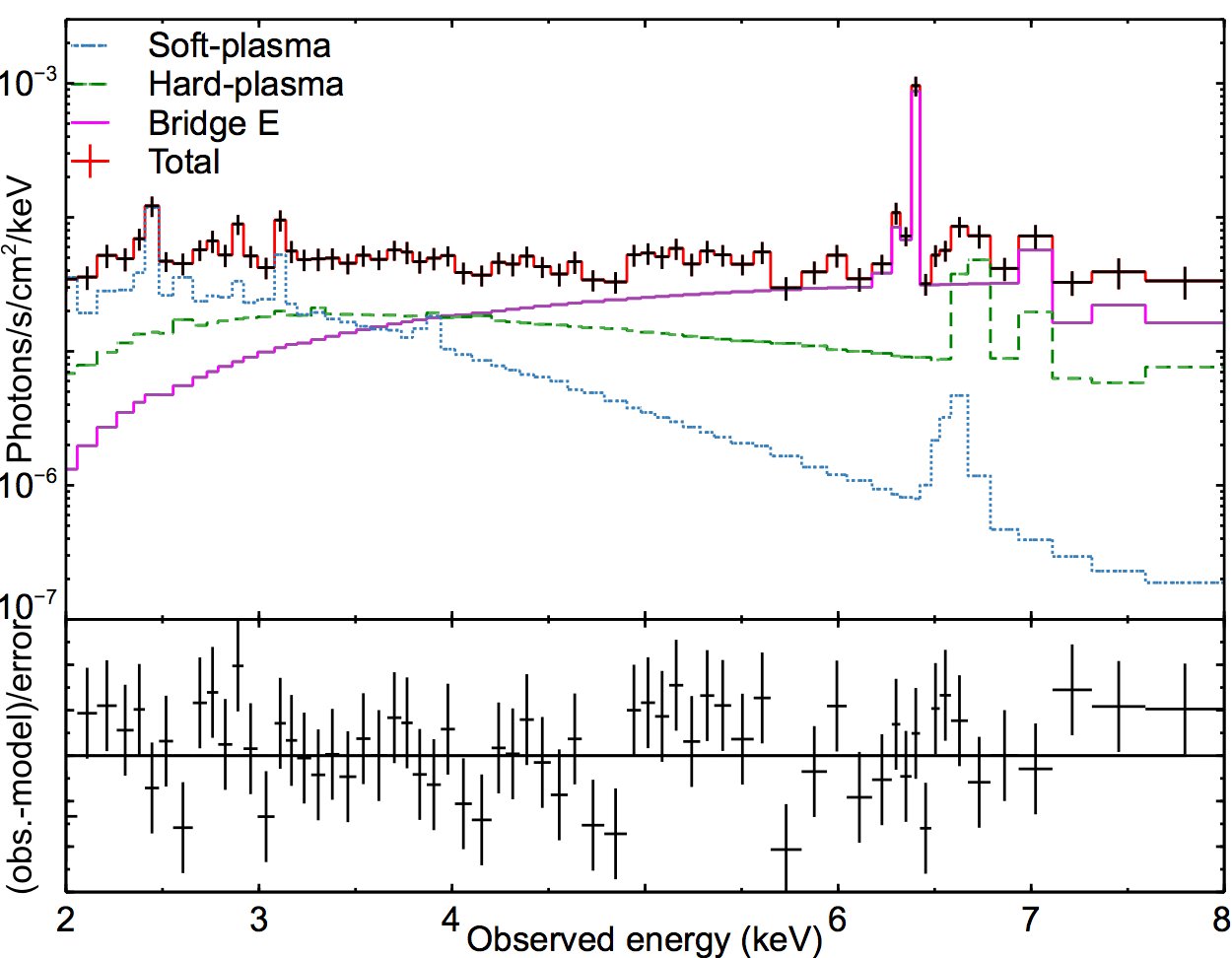}
\includegraphics[width=0.5\textwidth]{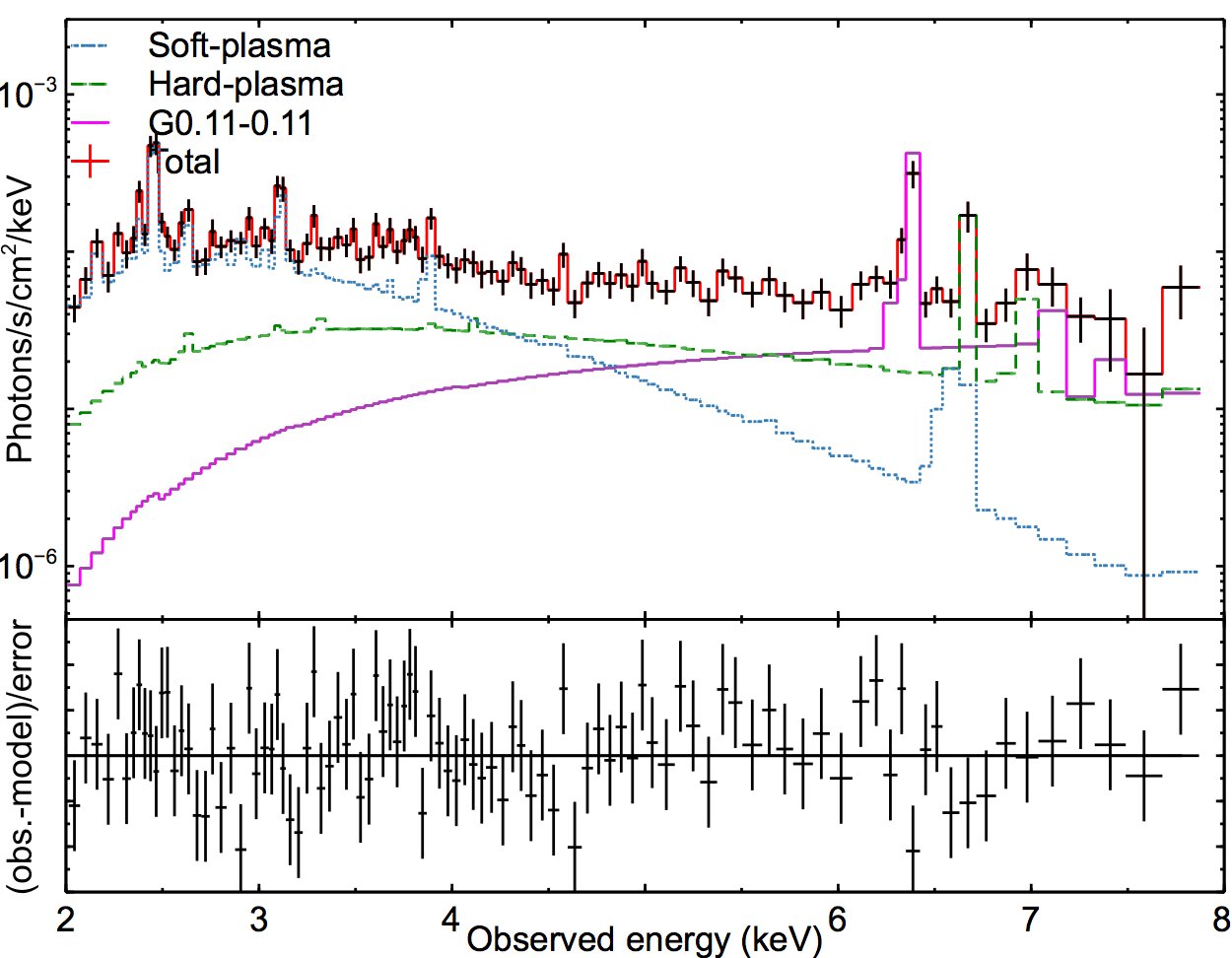}
 \includegraphics[width=0.5\textwidth]{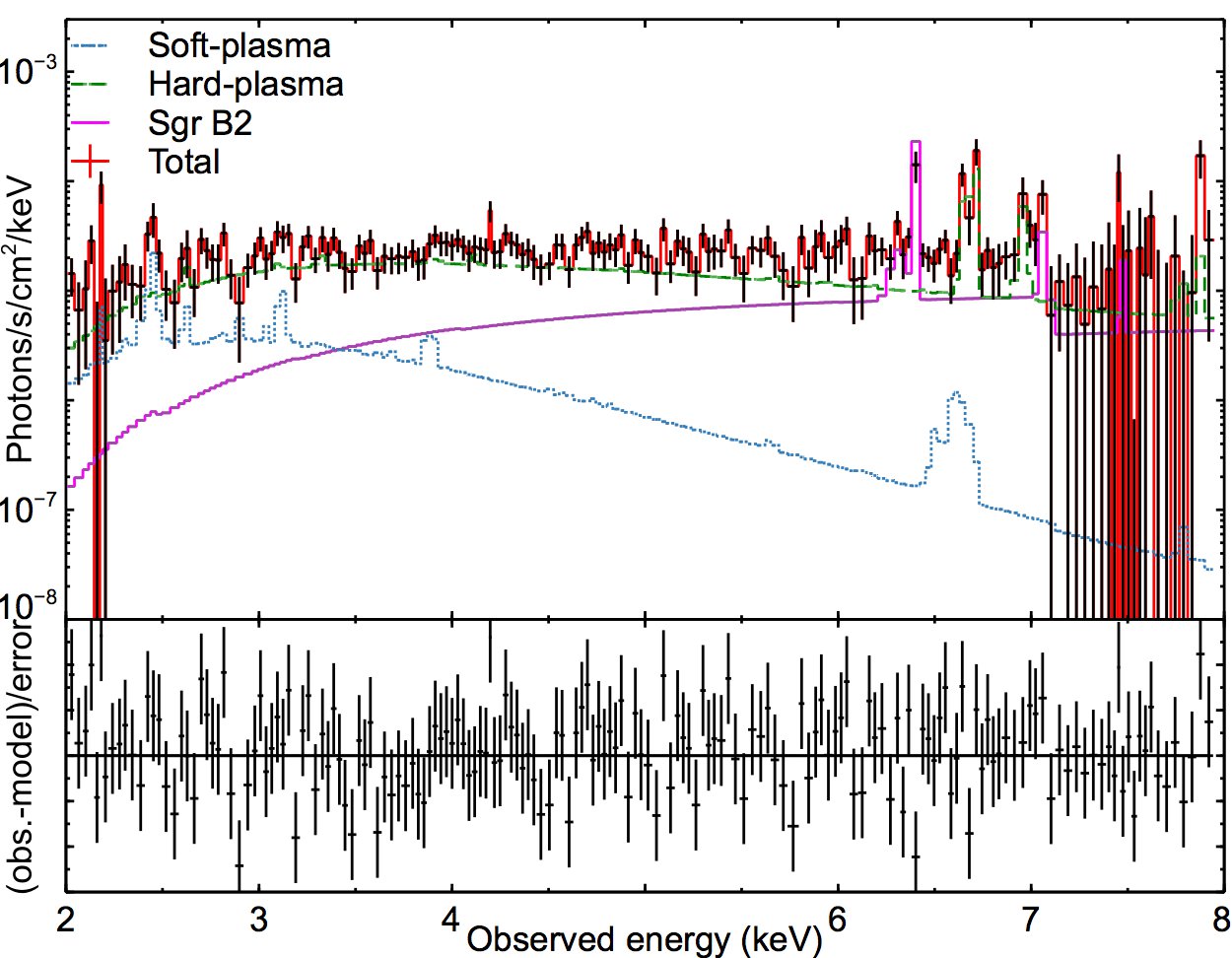}
 \includegraphics[width=0.5\textwidth]{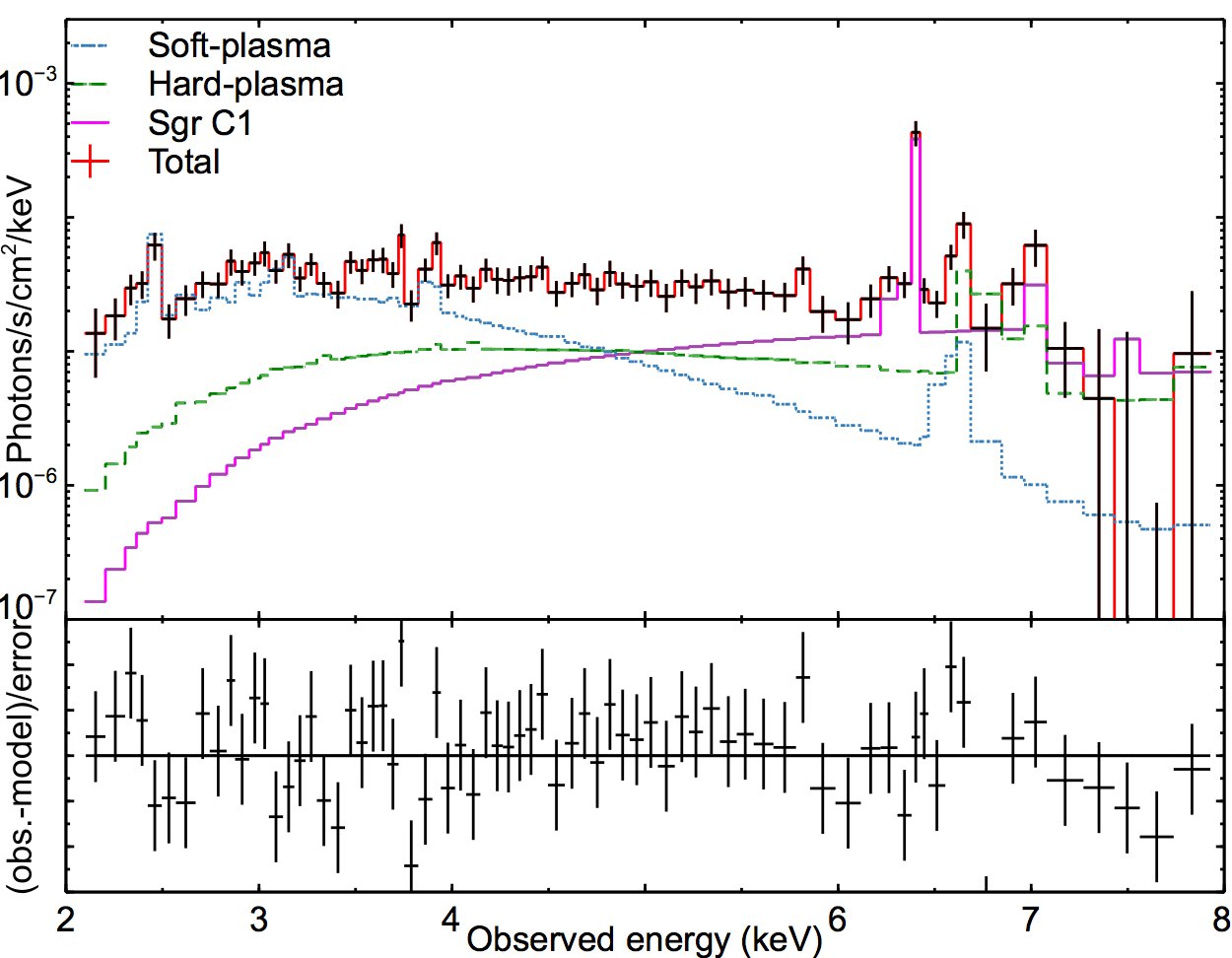}
 \includegraphics[width=0.5\textwidth]{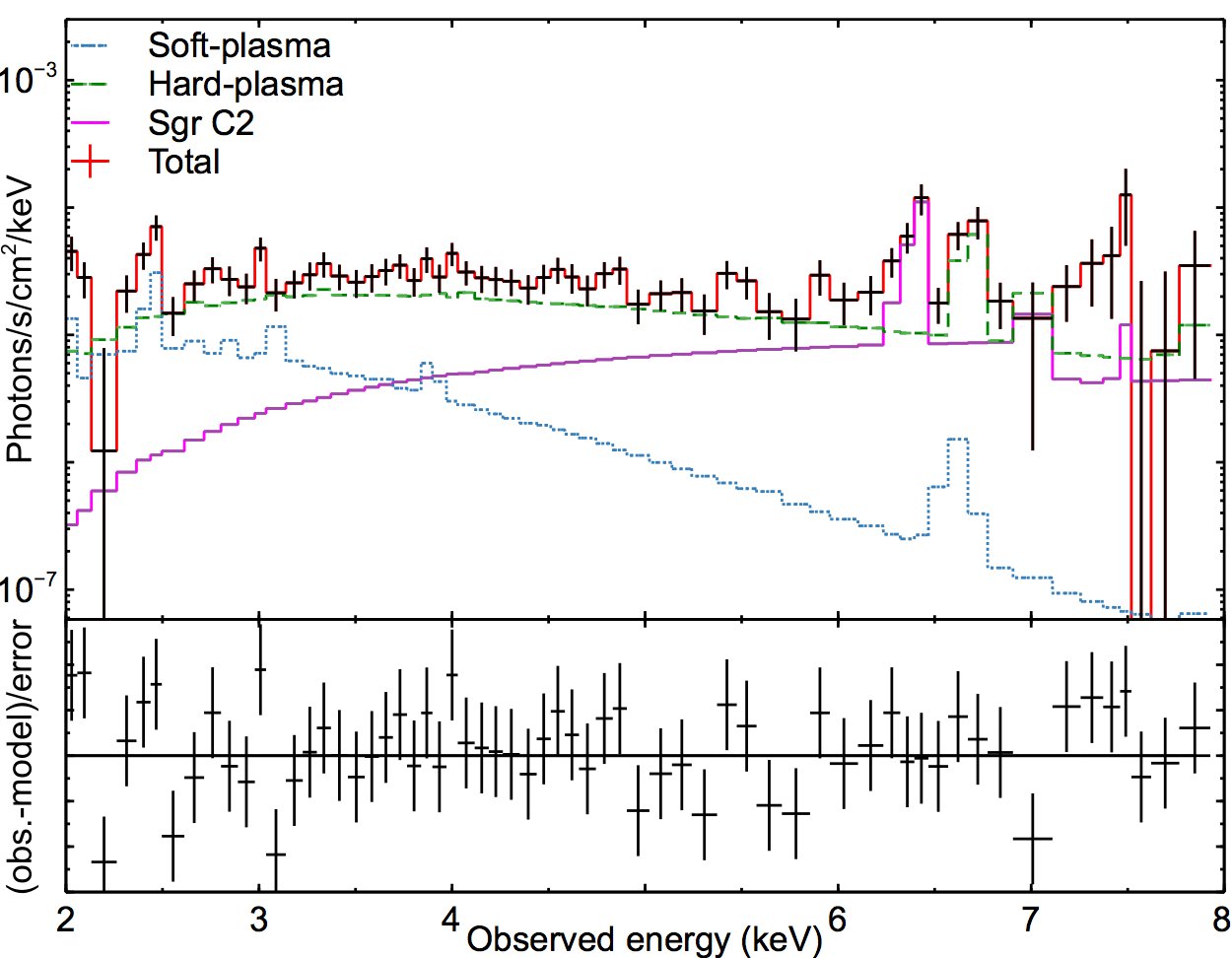}
 \includegraphics[width=0.5\textwidth]{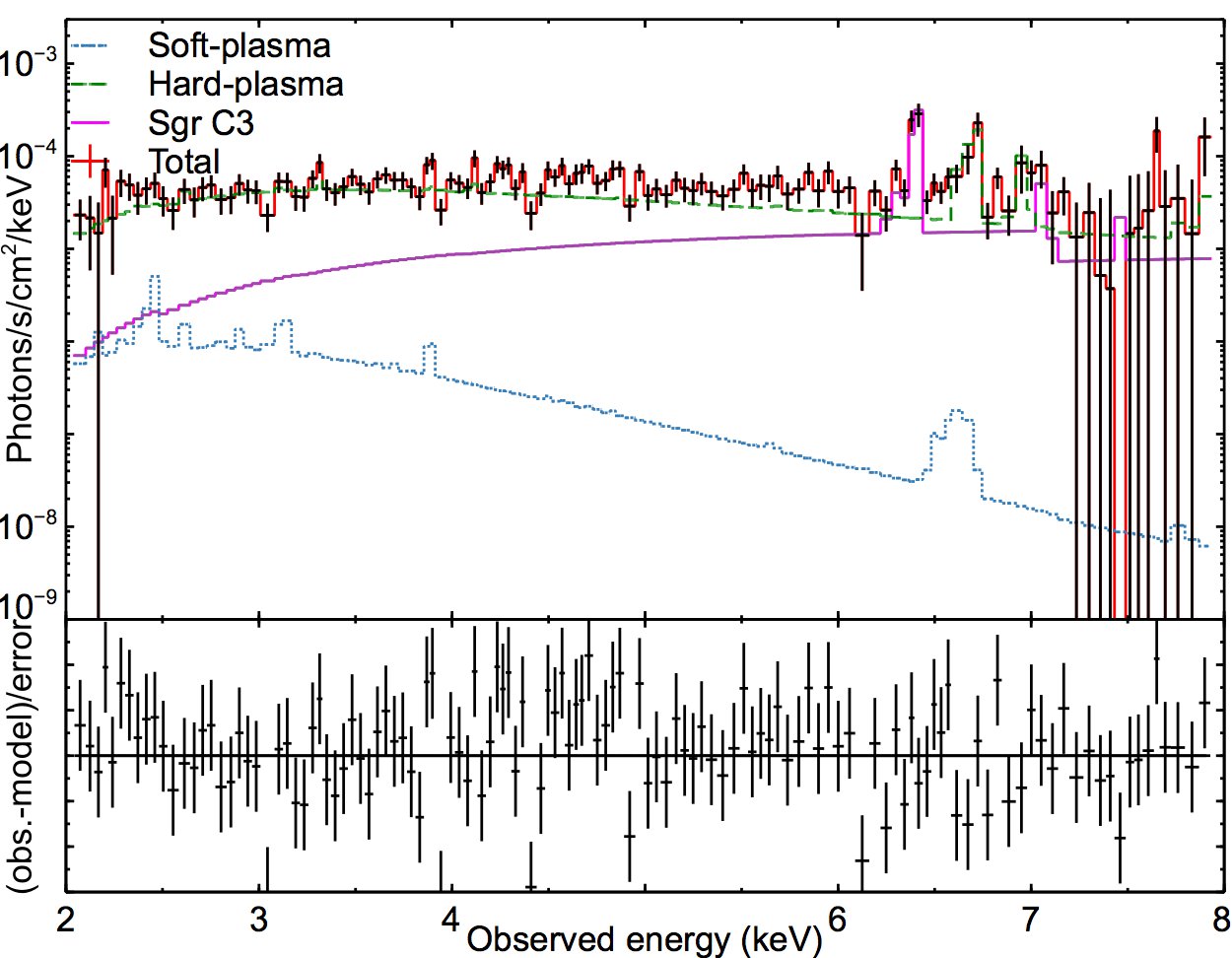}
\end{minipage}
 \caption{From top to bottom: Unfolded spectra and residuals to the best-fit model
 for MC2, Bridge B2, Bridge E, G0.11-0.11, Sgr B2, Sgr C1, Sgr C2, and Sgr C3.
 The total best-fit model and the reflection component are displayed as a solid line.
  The spectrum of the hard plasma is displayed as a dashed line. The spectrum
  of the soft plasma is displayed as a dotted line. }
  \label{allspec.fig}
\end{figure*}

%
\begin{table*}
\caption{Results of the spectral analysis of the molecular clouds described in Sect. \ref{chspec}.}     
\label{input2.tab}      
\centering                    
\begin{tabular}{lcc}        
\hline\hline                 
Target
& \nh \tablefootmark{a}
& Model component fluxes\tablefootmark{b}\\ 
\hline
%
%
&\multirow{3}{*}{} 
& Soft plasma: 2.0-4.0 keV \quad 4.0-8.0 keV\\
& & Hard plasma: 2.0-4.0 keV \quad 4.0-8.0 keV \\
& & Cloud: 2.0-4.0 keV \quad 4.0-8.0 keV  \\
%
%
& ($10^{22}$ \colc)
& ($10^{-13}$ \ergsc)\\
\hline
%
\multirow{3}{*}{MC2} 
& \multirow{3}{*}{$6 \pm 4 $}
& $2 \pm 1$ \quad $5 \pm 4$\\
&  & $1.3 \pm 0.2$ \quad $3.9\pm 0.8$  \\
&   & $0.13 \pm0.06 $ \quad $1.9 \pm 0.6 $  \\
\hline
%
%
\multirow{3}{*}{Bridge B2} 
& \multirow{3}{*}{ $\leq 8 $}
& $2.2 \pm 1.4 $ \quad $0.3 \pm 0.2$\\
&  & $1.7\pm 0.5$ \quad $1.8 \pm 0.8$  \\
&  & $0.5 \pm 0-9 $ \quad $4.7 \pm 0.8 $  \\
\hline
%
%
\multirow{3}{*}{Bridge E} 
& \multirow{3}{*}{ $4 \pm 1 $}
& $2.0 \pm 0.9 $ \quad $0.5 \pm 0.2$\\
&  & $1.9\pm 0.3$ \quad $4.7 \pm 0.8$  \\
&  & $1.19 \pm 0.09 $ \quad $14 \pm 1 $  \\
\hline
%
%
\multirow{3}{*}{G0.11-0.11} 
& \multirow{3}{*}{$7.0 \pm 0.3$}
& $8 \pm 1 $ \quad $3.2 \pm 0.4$\\
& & $2.7 \pm 0.4 $ \quad $9 \pm 1 $  \\
& & $0.71 \pm 0.08 $ \quad $11 \pm 1 $  \\
\hline
%
%
\multirow{3}{*}{Sgr B2} 
& \multirow{3}{*}{$8\pm2$}
& $0.4 \pm 0.3$ \quad $0.2 \pm 0.1 $\\
& & $1.4 \pm 0.1 $ \quad $5.4 \pm 0.5 $  \\
& & $0.21 \pm 0.03$ \quad $3.7 \pm 0.4$  \\
\hline
%
\multirow{3}{*}{Sgr C1} 
& \multirow{3}{*}{$12\pm1$}
& $2.5\pm 0.7$ \quad $ 1.8 \pm 0.5$\\
& & $0.6 \pm 0.2$ \quad $4 \pm 1 $ \\
& & $0.25 \pm0.03 $t \quad $6.1 \pm 0.8 $ \\
\hline
%
%
\multirow{3}{*}{Sgr C2} 
& \multirow{3}{*}{$7 \pm 2$}
& $0.7 \leq 0.4$ \quad $0.3 \pm 0.2$ \\
& & $1.7  \pm 0.2$    \quad $5.6 \pm 0.7$  \\
& & $0.26 \pm 0.05$ \quad $3.9 \pm 0.7 $  \\
\hline
%
%
\multirow{3}{*}{Sgr C3} 
& \multirow{3}{*}{$7 \pm 2$}
& $0.09 \pm 0.05$ \quad $0.03 \pm 0.02$ \\
& & $3.7  \pm 0.4$    \quad $12 \pm 1$  \\
& & $0.46 \pm 0.07$ \quad $8 \pm 1 $  \\
\hline
\end{tabular}
\tablefoot{
\tablefoottext{a}{Galactic hydrogen column density.}
\tablefoottext{b}{Fluxes of each model component in the quoted bands.}}
\end{table*}

\subsection{Chandra spectral analysis}
\label{chspec}
The last necessary
ingredient for simulating
IXPE observations of the
selected targets is the
spectral shape of each emission
component.
For all the regions listed in Table
\ref{input.tab},
we extracted the
spectrum from the most recent
available Chandra observation.
These are highlighted in bold
in Table \ref{obs.tab}.
We confirmed that
the extraction regions 
include no contamination 
of known
bright X-ray sources 
(listed in, e.g., \citealt{terrier2018}).\\
To extract the spectra, we used
the CIAO script specxtract,
which creates the source
and background spectra
and the necessary weighted response
matrices. We used the
customized blank-sky event file
to extract the background
spectrum in the same region.
We binned the spectra requiring
that a minimum of 30 counts 
is reached in each spectral bin.\\
We fit all the spectra 
in the 2.0-8.0 keV band with
Xspec version 12.10.1. We used 
a model including the Galactic
absorption,
the soft and hard plasma,
and the cloud emission. For the Galactic
absorption we used the phabs model,
with the hydrogen column density
\nh as a free parameter. For
the plasma components, we used
a collisionally ionized plasma
model \citep[APEC,][]{smith2001} 
with a temperature
set to 1.0 and 6.5 keV
for the soft and hard plasma,
respectively.
We considered solar
abundances and set the
redshift to zero.
For the molecular clouds, 
we used the neutral reflection
PEXMON model  \citep{nandra2007}, where we set
(as in, e.g., \citealt{ponti2010})
the photon index $\Gamma$ to 2,
the disk inclination to 60\degr,
and the cutoff energy to 150 keV.
The free parameters
of our fits are therefore the Galactic \nh
and the normalization of each emission
component. 
We show the
spectra of all the clouds
in Fig. \ref{allspec.fig}.
We list the parameters and
errors resulting from our
spectral analysis
in Table \ref{input2.tab}.
All the spectral fits are
statistically acceptable 
(\chidof $\leq 1.3$).\\

\section{Simulation of IXPE observations}
\label{stra}
We simulated
IXPE observations of the targets listed
in Table \ref{input.tab} using the dedicated
simulation framework \textit{ixpeobssim}
\citep{pesce2019}. This is a python-based
tool that can be fed by an arbitrary 
source model, including morphological, 
temporal, spectral, and 
polarimetric information. 
Hence, the framework 
uses the  IXPE instrument response functions
(i.e., the PSF and the detector effective area) 
to produce the IXPE-simulated event files.
These can be used to create images, spectra,
and modulation curves in different bands.\\
For each target, we performed
the simulation in the
region listed in Table \ref{input.tab}
and centered the field of view
on the coordinates of the target.
Within the regions of interest, we simulated
all the components that contribute
to the diffuse X-ray emission. 
In addition to the polarized emission
of the molecular clouds, we thus included the soft and hard plasma, the cosmic X-ray background,
and the instrumental background in our simulations.
For each emission component,
we input in the simulation the
spectrum, the polarization degree, the polarization
position angle, and
when possible, the spatial morphology.
We took the polarization degree and
polarization angle of each
molecular cloud from 
the model of \citet{marin2015},
as listed in Table \ref{input.tab}.
We considered a polarization degree
that is constant with energy, but
null at the energy of the fluorescence
\fek \, line ($6.32-6.48$ keV).
The fluorescent lines from
spherically symmetrical 
orbitals are unpolarized.
Conversely, for the plasma components,
we considered a null polarization.
In the case of MC2, Bridge-B2,
Bridge-E, and G0.11-0.11
we were able to input in the simulator
the real morphology
of the hard and soft plasma
and of the clouds using
the Chandra maps described
in Sect. \ref{chtarget}.
In the case of the clouds in the Sgr B
and Sgr C region, the Chandra data
quality did not allow us to compute
separate maps of each emission component.
We therefore assumed a uniform 
morphology of all the components in the
region of interest for these clouds.\\
For both the instrumental and the sky background
we simulated a null polarization. The internal polarization of the
detector is below 1\% and thus
negligible. For the instrumental background,
we took the spectrum from the
measurement of the non-X-ray background
of the neon-filled detector that flew
on board of OSO-8 \citep{bunner1987}. 
The gas mixture and absorption
coefficient of the OSO-8 detector
were similar to the one of the IXPE GPD.
For the instrumental
background, we simulated a uniform morphology on
the detector. In the simulation, the instrumental
background is internal to the detector, thus
it is not convolved with the instrumental
response functions. Finally,
for the sky background, we used the parameters
of the CXB spectrum of \citet{moretti2009}, 
and we renormalized it
to match the simulated sky area. We simulated
it as a sky source with a uniform morphology.
%
%
%

\begin{table*}
        \caption{Simulations results for the reflection nebulae.}     
        \label{output.tab}      
        \begin{tabular}{lcccc}        
                \hline\hline                 
                Target
                & Scaled P  \tablefootmark{a}
                & Diluted P  \tablefootmark{b}
                & MDP (2 Ms) \tablefootmark{c}
                & $F_{\rm min}$ \tablefootmark{d}\\
                \hline
                %
                %
                & 2.0-4.0 keV\quad 4.0-8.0 keV 
                & 2.0-4.0 keV \quad 4.0-8.0 keV 
                & 2.0-4.0 keV \quad 4.0-8.0 keV  
                &2.0-8.0 keV \\
                & (\%) 
                & (\%) 
                & (\%) 
                & ($10^{-13}$\ergsc) \\
                \hline
                MC2 \tablefootmark{*}
                & 0.8\%$-$1.6\% \quad 5\%$-$10\%
                &$ \leq 1\% $ \quad $ 5\%$
                & 15\% \quad 19\%
                & 0.2\\
                Bridge B2 \tablefootmark{*}
                & 1.9\%$-$2.7\% \quad 9\%$-$12\%
                         &$3 \% $ \quad $8 \%$
                & 14\% \quad 20\%
                &0.1  \\
                %
                %
                Bridge E \tablefootmark{*}
                & 2.6\%$-$3.1\% \quad 8.5\%$-$9.9\%
                         &$3 \% $ \quad $7 \%$
                & 11\% \quad 12\%
                        & 0.3 \\
                G0.11-0.11 \tablefootmark{*}
                & 3.1\%$-$3.9\% \quad 23\%$-$29\%
                &$3 \%$ \quad $16 \%$
                & 7\% \quad 9\%
                &0.5 \\
                \hline
                %
                Sgr B2 \tablefootmark{**}
                & 6\%$-$8\% \quad 23\%$-$29\%
                 &$13 \%$ \quad $26 \%$
                & 26\% \quad 21\%
                & 3.5  \\
                \hline
                Sgr C1\tablefootmark{**}
                &3.5\%$-$4.6\% \quad 18\%$-$23\%
                &$1 \%$ \quad $10 \%$ 
                & 13\% \quad 14\%
                & 0.7 \\
                Sgr C2 \tablefootmark{**}
                & 4\%$-$6\% \quad 12\%$-$27\%
                & $4 \%$ \quad $10 \%$ 
                & 15\% \quad 15\%
                & 1.1 \\
                Sgr C3 \tablefootmark{**}
                & 3\%$-$4\% \quad 10\%$-$14\%
                & $3 \%$ \quad $8 \%$ 
                & 12\% \quad 11\%
                & 2.3 \\
                \hline
        \end{tabular}
        \tablefoot{
                \tablefoottext{a}{Obtained from the fluxes and errors listed in Table \ref{input.tab}.}
                \tablefoottext{b}{Obtained from mock simulations reaching an MDP of 1\%.
                By design, the absolute error on the diluted polarization degree is of 1\% or lower.}
                \tablefoottext{c}{Obtained for 2 Ms exposure time.}
                \tablefoottext{d}{Minimum flux detectable by IXPE in 2 Ms with a signal-to-noise ratio of at least 3.}
                \tablefoottext{*}{Simulation performed using Chandra maps to define the morphology of all the components.}
                \tablefoottext{**}{Simulation performed assuming a uniform morphology for all the components.}}

\end{table*}
%

%
%
\section{Results and discussion}
\label{disc}

\begin{figure}
 \includegraphics[width=0.5\textwidth]{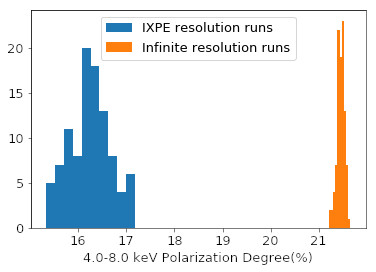}
 \caption{Histograms showing the distribution of the polarization degree in
  the 4.0-8.0 keV band obtained
  by simulating the cloud G0.11-0.11 for different instrumental resolutions. 
  The orange histogram shows the case with infinite spatial resolution.
The blue histogram represents the case with IXPE resolution.}
  \label{dilution.fig}
\end{figure}
%
%
%
\begin{figure*}
\begin{minipage}[c]{1.0\textwidth}
\includegraphics[width=0.5\textwidth]{g011_map_arrow.png}
\hspace{10pt}
 \includegraphics[width=0.5\textwidth]{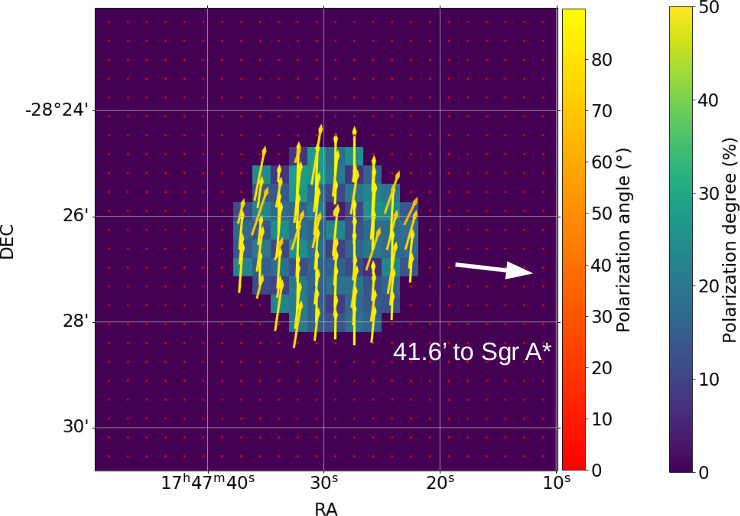}
 
\end{minipage}
 \caption{Simulated IXPE polarization maps of G0.11-0.11 (left panel) 
 and Sgr B2 (right panel). The background is color-scaled according
 to the polarization degree. The colored arrows represent the direction of the polarization
 angle and are color-scaled accordingly. The color scales for the polarization degree
 and angle are shown at the right of each figure. The direction of Sgr A* is
 also indicated for comparison.}
  \label{polmap.fig}
\end{figure*}

%
%
%
Using the input
ingredients described
in Sect. \ref{ch}
and the procedure described
in Sect. \ref{stra}, we simulated 
IXPE observations of all the
targets. We extracted two main
quantities from
the simulations: the degree to which polarization is diluted by
the ambient and background
radiation, and which MDP
can be reached in a realistic 
exposure time. These
pieces of information
serve to evaluate the
detectability of the considered
targets in an X-ray polarimetric
study of the GC.\\
%
%
In order to
obtain a sensible
measurement of the
diluted
polarization degree, 
we proceeded
as follows. For all
the targets, we ran mock 
simulations of observations
reaching an MDP of at least 1\%.
Thus, the mock exposure
time (i.e., 100 Ms) was chosen to obtain
that the absolute error
on the polarization degree
is 1\% or lower.
This mimics
an ideal case
where the statistical uncertainty
of the determined polarization
degree is negligible. Any observed difference
between the determined
polarization degree
and the theoretical one
in these simulations must thus be caused
by the mixing between
polarized and unpolarized
components. We
note  indeed that in
simulations
without
unpolarized sources in the field
of view, the theoretical 
polarization degree
is always recovered within 3\% or less when the
MDP of the simulation is
at least 1\%. \\
In 
Table \ref{output.tab}
we list the diluted
polarization degrees
resulting from the simulations
and compare them
with the scaled
polarization degrees that
result from a simple
rescaling using the ratio
between the reflection flux
and the total flux 
(e.g., \citealt{marin2015}).
We consider that
the scaled polarization
degrees are affected by the uncertainty
of the spectral decomposition.
The ranges given in Table \ref{output.tab}
are obtained as $P \times (F_{\rm cloud} \pm eF_{\rm cloud})/F_{\rm tot} $
, where $F_{\rm cloud}$ and $eF_{\rm cloud}$
are the flux and error, respectively,
for the cloud component, $F_{\rm tot}$
is the total flux  and $P$ is
the theoretical polarization degree.
We observe
that the diluted polarization degrees
are in some cases lower
than the scaled polarization degrees.
This additional dilution 
must be induced by
the morphological smearing 
of the source due to the finite
PSF. We illustrate this
point in Fig. \ref{dilution.fig}.
We ran 100 simulations of G0.11-0.11
for an ideal case of an instrument
with infinite spatial resolution
and zero background
and 100 normal 
simulations, where
the convolution with the
instrumental PSF is
considered. 
In this
exercise, we considered
a mock exposure
time of 100 Ms, so that
the statistical fluctuations
of the simulated
polarization degree
were within 1\%.
In Fig. \ref{dilution.fig}
we compare the distribution
of the polarization
degree obtained in the
two cases.
We found that an
instrument with infinite
spatial resolution would
observe a polarization
degree of $\sim$21\%,
consistent with what
is predicted by a
simple rescaling
of the flux. An
instrument with
the spatial resolution
of IXPE would observe
an additional dilution of $\sim$ 5\%.
This difference is not
explained by the statistical
fluctuations of the simulation result because
that is by design
lower than 1\% in our
simulations.
In conclusion,
our work shows that the
finite spatial
resolution of the
polarimeter can
add a sensible
additional dilution
depending on
the extension
and on the 
morphological
details of the
source. The quality
of the imaging output
plays a significant role
for an X-ray
polarimetric study
of the GC
region, where
the polarized regions
have to be resolved out
of the surrounding unpolarized 
emission.\\
%
%
The diluted polarization degrees
have to be compared
with the MDP that can be achieved
in a realistic exposure
time. From our IXPE
simulations,
we computed the MDP
in the 2.0$-$4.0 keV
and
4.0$-$8.0 keV band
by running realistic
simulations with an exposure
time of 2 Ms. 
We note
that polarimetry
is a photon-starving
science and $\text{an}$
 exposure time
of some milliseconds may be required
for faint or
lowly polarized
sources (e.g., for
extragalactic sources
such as AGN). Even
for bright Galactic
sources or extragalactic
blazars exposure times
of some hundred
kiloseconds are typically required.\\
From the MDP listed in
Table \ref{output.tab}
we can derive a first indication of the
preferable targets for IXPE. We
found that the most suitable energy
band for searching for polarization
signatures is the 4.0$-$8.0 keV
band, where the emission
of the molecular clouds
dominates the flux output.
This exercise
indicates that the most promising
targets for IXPE observation
are G0.11-0.11 and Sgr B2.
For these two targets,
we found that the diluted
polarization degree in
the 4.0$-$8.0 keV band
is higher than the 
MDP that can be reached in a 2 Ms
long IXPE observation.
Our simulations therefore confirm
the preferable targets that were
previously suggested by
\citet{marin2015}.\\
However, some caveats must be considered in
the planning of an X-ray polarimetric
study of the GC. The
first issue that we investigated
is that the 
flux of the molecular clouds
varies with time.
The flux levels we considered
are those of 2017 for MC2, Bridge B2, Bridge E,
and G0.11-0.11; of 2010 for Sgr B2;
of 2014 for Sgr C1 and
Sgr C2; and of 2007
for Sgr C3. Our simulations
indicate that at these
flux levels, an IXPE observation
of any of these targets will always
be source dominated. For
instance, we find that
for the faintest target
of the pools considered
here (i.e., Sgr C3), the
instrumental background
accounts for 2\%
of the total counts, while
the CXB accounts for
3\% of the total counts.
Nonetheless, by the time
of the IXPE observation,
the flux of the molecular clouds
may be higher or lower
than we considered here.
In a recent study
of the long-term flux variability
of the molecular clouds,
\citet{terrier2018}
found that  MC2,
G0.11-0.11, and Sgr B are
fading while the Bridge is
brightening up. The trend
for Sgr C is more stable,
although within a larger
uncertainty.
It is
therefore useful 
to compute the minimum
flux that would be detectable by IXPE
in 2 Ms for each target with a signal-to-noise ratio of
at least 3. 
Exploiting our
estimates of the background 
contribution,
we determined
these flux thresholds
and list them in
Table \ref{output.tab}.
We found
that the targets in the Sgr A
field remain detectable
unless the total 
flux decreases by one
(e.g., for MC 2 and Bridge B2)
or even two orders of magnitude
(e.g., for Bridge E and
G0.11-0.11) with  respect
to the level we considered here.
In the case
of Sgr B2, the total flux
would need to be lower by a factor
3 with respect
to the level observed
in 2010 (i.e., 1.1 $\times 10^{-12}$
\ergsc) to fall below
the detection threshold.\\
%
In addition to the
variability in flux, the
molecular clouds in
the GC also exhibit
variability in morphology. 
For instance, the brightest
centroid in Sgr C2 underwent a displacement
of 1.6\arcmin \,
in 12 years \citep{terrier2018}. 
We investigated
the effect of the morphology
for the result of our simulations.
At first, we assessed 
the effect of positioning 
the simulated IXPE pointing
well onto the brightest \fek \th patch.
We tested this issue using the
2 Ms long simulation
of the Bridge-B2 cloud, which
displays a well-defined bright
knot. We find
that shifting the
IXPE pointing
just $\sim$20 \arcsec
away from the brightest
patch causes a loss
of $\sim$300 counts
and decreases
the MDP by 1\%.
This suggests that
it is convenient
to center the IXPE
pointing on a bright knot
in order
to maximize the collected counts
and thus the chance
of detecting
a significant
polarization.\\
We therefore
evaluated the effect
of the morphology
on determining
the diluted polarization
in the region of interest.
In figure \ref{polmap.fig}
we show as an example the
simulated IXPE polarization maps 
of the two best targets. 
These were produced from
the mock simulations.
In these maps, the colored arrows
indicate the direction
of the polarization angle.
In the case of a reflection
nebula, this is normal to the 
projected direction
of the illuminating source. 
In the simulated map of
Sgr B2, the nebula is
uniform in color and polarization
degree because it was simulated
assuming a uniform morphology
for all the components.
In the simulated map
of G0.11-0.11, this was
obtained by starting from
the Chandra maps of
the different components,
and the irregular distribution
of polarization fraction and color
within the nebula reflects
the different level of mixing
between polarized and unpolarized
emission. \\
Nonetheless,
the dilution of the polarization
degree averaged over
the regions of interest
mildly depends on the internal
morphology, likely
because the
substructures are 
one scale smaller
than the IXPE PSF. 
We verified this
point by running
simulations of the G0.11-0.11
field assuming 
a uniform morphology
for all the components
and a mock exposure
time of 100 Ms.
The results for
the diluted polarization
degree are the same
within the uncertainty
as in the run using
the Chandra maps.
We are therefore a posteriori confident
that our estimates
of the polarization
dilution in Sgr B2,  Sgr C1,
Sgr C2, and Sgr C3              
are trustworthy.\\
All in all, we remark
that an X-ray observation
of the GC would be useful
prior  the IXPE pointing.
With the Spectrum R{\"o}ntgen
Gamma (SRG) on board of eROSITA, for example,
it is possible to measure the flux level
of the candidate targets. With
Chandra or XMM-Newton, it
is possible to determine
which patches are currently
illuminated. This would
help in deciding the
best pointing.\\
%
%
Finally, in Table
\ref{dist.tab} we investigate
the most critical
uncertainty that
affects the evaluation
of the detectability
of the polarization
of the molecular cloud.
The theoretical
polarization degree
relies on the poorly
constrained line-of-sight 
distance of the cloud
and will be corrected
when a more
robust determination
of $d_{\rm los}$ is found.
We searched in the literature
for determinations of the
line-of-sight distance of the
clouds different from
those assumed in \citet{marin2015}
(listed as 
$d_{\rm los}^{\rm other}$ in Table
\ref{dist.tab}).
These were obtained in works
where the scattering
angle is derived
from modeling
of the reflection
spectrum 
\citep{capelli2012, walls2016, chuard2018}
and were often loosely
constrained.
Starting from
the range of  $d_{\rm los}^{\rm other}$,
we used equations
\ref{eqp} and \ref{eqd}
to compute 
the correspondent range
of polarization degree ($P^{\rm other}$),
and we used the dilution factors
in the 4.0-8.0 keV band 
that can be inferred from
Table \ref{output.tab} to
determine the corresponding
range in diluted polarization degree
 ($P^{\rm other}_{\rm dil}$).
Thus, we were able
to verify whether for
a different assumption
on $d_{\rm los}$, 
the diluted polarization
degree of our targets rises or drops below the MDP
that can be obtained by IXPE in
the 4.0-8.0 keV band
in 2 Ms.
The values listed
in Table \ref{dist.tab}
confirm
the detectability
of G011-0.11 and Sgr B2
also for other possible distances
reported in the literature.
The molecular clouds
Bridge B2, Bridge E,
and Sgr C1 might 
be detectable if their
real distance along the line
of sight lies within the
upper bound of the
range determined by \citet{capelli2012}
and \citet{chuard2018}.\\
We also investigated how
the enhanced sensitivity
of eXTP allows enlarging
the pool of suitable targets.
The effective area
of eXTP will
be larger  
by a factor $\sim$4,
which implies
(using
equation \ref{mdp})
that the MDP for the case
of eXTP is lower
than those of IXPE
by a factor 0.51.
When
this factor is applied to the MDP values listed in Table \ref{output.tab}, this
implies that
G0.11-0.11,
Sgr B2, Sgr C1, Sgr C2,
and Sgr C3
are potential
targets for eXTP
in the 4.0-8.0 keV
band. The ranges
of diluted
polarization degrees
obtained in Table \ref{dist.tab}
by relaxing the constraints
on $d_{\rm los}$
offer a window
of eXTP detectability
for virtually all the targets.
More sensitive
telescopes, for instance,
the X-ray Polarimetry Probe 
\citep[XPP,][]{xpp} 
or
the New Generation X-ray Polarimeter 
\citep[NGXRP,][]{esa2050}
mission concept
would allow detecting the X-ray polarization of
the molecular clouds
with shorter exposure times.\\
In conclusion, an X-ray polarimetric
study of the CMZ is a challenging
experiment because of the
dynamic behavior of the
reflection emission and because
of the complex gaseous environment
in which the nebulae are embedded.
In this work,
we set up a simulation method
that allows realistically
assessing how some 
critical factors
(i.e., the variability in flux
and morphology of the clouds,
and the dilution of the polarization degree in
the unpolarized ambient and background radiation) 
affect the detectability of a reflection
nebula observed on
axis. Nonetheless, other levels of complexity
remain unexplored. In a future expansion
of this work, we will produce
a simulated IXPE map
of the entire Sgr A 
field of view. This would allow
us to investigate, for instance,
how the detectability degrades
for a nebula off axis and what happens
in regions where
gas filaments
with a different level of
polarization are mixed.
Because the time required to make
a significant measurement
of the reflection nebulae in the GC
is some$\text{ milliseconds}$,
the impact on the planning
of IXPE observations is
significant. Our realistic predictions
are therefore important to inform
the decision
of including these observations
in the planning.\\

%
%
%
\begin{table}
\caption{Polarization obtained for alternative values of
$d_{\rm los}$ reported in the literature.}     
\label{dist.tab}      
\centering                    
\begin{tabular}{lcccc}        
\hline\hline                 
Target
& $d_{\rm los}^{\rm other}$ \tablefootmark{a} 
& $P^{\rm other}$  \tablefootmark{b} 
& $P^{\rm other}_{\rm dil}$  \tablefootmark{c} 
& Ref.  \tablefootmark{d}\\
%
& (pc)
& (\%)
& (\%) 
& \\
\hline
%
%
MC2
& -29.7$-$7.3
&50$-$53
&9$-$10
& A \\
%
%
Bridge B2
& -6.9$-$6.9
& $\leq 84$
& $\leq 42$
& A \\
%
%
Bridge E
& -13.7$-$13.7
& $\leq 83$
& $\leq 45$
& A \\
%
%
G0.11-0.11
& -3.1$-$3.1
& $\leq 93$
& $\leq 26$
& A \\
%
%
Sgr B2
& -50$-$-47
&  61$-$83
&  24$-$33
& B \\
%
%
Sgr C1
& -0.61$-$47
&  50$-$99.9
&  16$-$32
& C \\
%
%
Sgr C2
& -38$-$-25
&  50$-$54
&  14$-$16
& C \\
\hline
\end{tabular}
\tablefoot{
\tablefoottext{a}{Range of $d_{\rm los}$ from the quoted references.}
\tablefoottext{b}{Range of polarization degree corresponding to  $d_{\rm los}$, obtained from Eqs. \ref{eqp} and \ref{eqd}.}
\tablefoottext{c}{Range of diluted polarization degree obtained from the values of Table \ref{output.tab}}
\tablefoottext{d}{A: \citet{capelli2012}, B: \citet{walls2016}, and C: \citet{chuard2018}.}}
\end{table}


\section{Summary and conclusions}
\label{conc}
Measuring 
the X-ray polarization property
of a reflection nebula
in the GC allows
us to confirm (or discard)
that they are illuminated
by a past outburst  of Sgr A*
(through the polarization
angle) and 
to determine the position
of the nebula along
the line of sight
(through the polarization
degree). These
are critical uncertainties
that hamper our ability
of using the variability
of the reflection emission
to infer how our Galactic
nucleus was behaving
$\text{a }$few hundred years
ago. Assessing
the history of our
Galactic nucleus 
has implications
for our understanding
of the duty cycle of
mass accretion onto
SMBH that is believed
to drive to the coevolution
of SMBH and galaxies.\\
We have evaluated
the feasibility
of this experiment
with IXPE, which is expected
to launch in 2021,
and
with eXTP, which is
scheduled for launch in 2027. We simulated IXPE observations
of the molecular clouds MC2, Bridge-B2, 
Bridge E, G0.11-0.11, Sgr B2,
Sgr C1, Sgr C2 and, Sgr C3 considering
the polarization properties
predicted by the model
of \citet{marin2015}.
We used the Monte Carlo-based
simulation tool \textit{ixpeobssim}
to individually simulate IXPE images of 
these targets.
In our simulations, we considered 
the spectrum (using Chandra spectra), 
the polarization properties,
and (when possible,
using Chandra images)
the spatial morphology
of the molecular clouds and
of the diffuse emission
that is comprised 
in the region of interest.
We modeled the diffuse emission
of the GC using two thermal
plasma components ($T_{\rm soft-plasma}\sim$ 1 keV
and $T_{\rm hard-plasma}\sim$ 6.5 keV).
Finally, we included in our simulations the
instrumental background and the cosmic
X-ray background. Our strategy
is designed to estimate
the degree to which the polarization degree
of the clouds is diluted by the
unpolarized ambient radiation and
by the morphological smearing
of the sources due to the
instrumental PSF.\\
We determined
for each cloud the minimum
flux that would be detectable
by IXPE in 2 Ms. We find
that the molecular clouds
considered here become
undetectable when  the total flux decreases
by a factor 3$-$100 (depending
on the cloud) with respect
to the level considered here.
Moreover, we found that the dilution
of the polarization degree
ranges between 0.3\% and 23\%
in the 2.0$-$4.0 keV
band and  19\% and 55\%
in the 4.0$-$8.0 keV band.
We note that the morphological
smearing of the sources
contributes additional dilution, whose
value varies from cloud to cloud.
The diluted polarization degree does not
depend on the internal
morphology of the
gas in the region of
interest.\\
For the flux levels we considered and
the polarization degrees computed
by  \citet{marin2015}, the most
promising targets for IXPE observations
are G0.11-0.11 and Sgr B2. For these
two cases, we found that the 4.0-8-0 keV polarization,
even after being diluted by the surrounding plasma,
is detectable by IXPE with a 2 Ms
observation. However, the theoretical
polarization degree strongly depends
on the assumed position of the
cloud along the line of
sight.  If the assumption
on the distance is relaxed
within the range reported
in the literature, a wider range
of possible polarization degrees
can be derived. If this
is the case, then also
Bridge-B2, Bridge-E,
and Sgr C1 might be detectable
by IXPE in 2 Ms. \\
Because its effective area is larger by a factor $\sim$4 with
the same exposure time, eXTP
will be able to detect the 4.0$-$8.0 keV
polarization degree
predicted by \citet{marin2015} 
of G0.11-0.11, Sgr B2, Sgr C1, 
 Sgr C2, and Sgr C3. When
a more relaxed constraint
on the distance along the line
of sight is considered, then
all the targets considered
here may be detectable
by eXTP.\\
%
%
%

%

\begin{acknowledgements}
The Italian contribution 
to the IXPE mission is supported 
by the Italian Space Agency 
through agreements ASI-INAF n.2017-12-H.0
and  ASI-INFN n.2017.13-H.0.
FM  acknowledges the support
from the Programme National des Hautes
Energies of CNRS/INSU with INP and IN2P3, 
co-funded by CEA and CNES.
We thank Gabriele Ponti
and Alessandra De Rosa
for useful chats about
the Chandra data analysis
and the eXTP capability.
We thank the anonymous referee
for the helpful comments 
that improved
this manuscript.
\end{acknowledgements}

\begin{appendix}
\section{Chandra analysis log}

\begin{table*}
\caption{Log of the Chandra observations.}     
\label{obs.tab}   
\centering                    
\begin{tabular}{cccccc}        
\hline\hline                 
Target
& Obs. ID
& Date 
& \multicolumn{2}{c}{Pointing}
& Exposure time\\
& & & Name & hh mm ss.s  & (ks)\\
\hline
MC2, Bridge-B2, Bridge E, G0.11-0.11 & 2951 &   2002-02-19 &  Sgr A$^{*}$ &  17 45 40.00  -29 00 28.10 & 12\\
MC2, Bridge-B2, Bridge E, G0.11-0.11 & 2952 &   2002-03-23 &  Sgr A$^{*}$ &  17 45 40.00  -29 00 28.10 & 12\\
MC2, Bridge-B2, Bridge E, G0.11-0.11 & 2953 &   2002-04-19 &  Sgr A$^{*}$ &  17 45 40.00  -29 00 28.10 & 12\\
MC2, Bridge-B2, Bridge E, G0.11-0.11 & 2954 &   2002-05-07 &  Sgr A$^{*}$ &  17 45 40.00  -29 00 28.10 & 12\\
MC2, Bridge-B2, Bridge E, G0.11-0.11 & 2943 &   2002-05-22 &  Sgr A$^{*}$ &  17 45 40.00  -29 00 28.10 & 38\\
MC2, Bridge-B2, Bridge E, G0.11-0.11 & 3663 &   2002-05-24 &  Sgr A$^{*}$ &  17 45 40.00  -29 00 28.10 & 38\\
MC2, Bridge-B2, Bridge E, G0.11-0.11 & 3392 &   2002-05-25 &  Sgr A$^{*}$ &  17 45 40.00  -29 00 28.10 & 170\\
MC2, Bridge-B2, Bridge E, G0.11-0.11 & 3393 &   2002-05-28 &  Sgr A$^{*}$ &  17 45 40.00  -29 00 28.10 & 158\\
MC2, Bridge-B2, Bridge E, G0.11-0.11 & 3665 &   2002-06-03 &  Sgr A$^{*}$ &  17 45 40.00  -29 00 28.10 & 90\\
\hline
MC2, Bridge-B2, Bridge E, G0.11-0.11 & 3549 &   2003-06-19 &  Sgr A$^{*}$ &  17 45 40.00  -29 00 28.00 & 25\\
\hline
MC2, Bridge-B2, Bridge E, G0.11-0.11 & 4683 &   2004-07-05 &  Sgr A$^{*}$ &  17 45 40.00  -29 00 28.00 & 50\\
MC2, Bridge-B2, Bridge E, G0.11-0.11 & 4684 &   2004-07-06 &  Sgr A$^{*}$ &  17 45 40.00  -29 00 28.00 & 50\\
\hline
MC2, Bridge-B2, Bridge E, G0.11-0.11 & 6113 &   2005-02-27 &  Sgr A$^{*}$ &  17 45 40.00  -29 00 28.00 & 5\\
MC2, Bridge-B2, Bridge E, G0.11-0.11 & 5950 &   2005-07-24 &  Sgr A$^{*}$ &  17 45 40.00  -29 00 28.00 & 48\\
MC2, Bridge-B2, Bridge E, G0.11-0.11 & 5951 &   2005-07-27 &  Sgr A$^{*}$ &  17 45 40.00  -29 00 28.00 & 49\\
MC2, Bridge-B2, Bridge E, G0.11-0.11 & 5952 &   2005-07-29 &  Sgr A$^{*}$ &  17 45 40.00  -29 00 28.00 & 45\\
MC2, Bridge-B2, Bridge E, G0.11-0.11 & 5953 &   2005-07-30 &  Sgr A$^{*}$ &  17 45 40.00  -29 00 28.00 & 49\\
MC2, Bridge-B2, Bridge E, G0.11-0.11 & 5954 &   2005-08-01 &  Sgr A$^{*}$ &  17 45 40.00  -29 00 28.00 & 18\\
\hline
MC2, Bridge-B2, Bridge E, G0.11-0.11 & 6639 &   2006-04-11 &  Sgr A$^{*}$ &  17 45 40.00  -29 00 28.00 & 5\\
MC2, Bridge-B2, Bridge E, G0.11-0.11 & 6640 &   2006-05-03 &  Sgr A$^{*}$ &  17 45 40.00  -29 00 28.00 & 5\\
MC2, Bridge-B2, Bridge E, G0.11-0.11 & 6641 &   2006-06-01 &  Sgr A$^{*}$ &  17 45 40.00  -29 00 28.00 & 5\\
MC2, Bridge-B2, Bridge E, G0.11-0.11 & 6642 &   2006-07-04 &  Sgr A$^{*}$ &  17 45 40.00  -29 00 28.00 & 5\\
MC2, Bridge-B2, Bridge E, G0.11-0.11 & 6363 &   2006-07-17 &  Sgr A$^{*}$ &  17 45 40.00  -29 00 28.00 & 30\\
MC2, Bridge-B2, Bridge E, G0.11-0.11 & 6643 &   2006-07-30 &  Sgr A$^{*}$ &  17 45 40.00  -29 00 28.00 & 5\\
MC2, Bridge-B2, Bridge E, G0.11-0.11 & 6644 &   2006-08-22 &  Sgr A$^{*}$ &  17 45 40.00  -29 00 28.00 & 5\\
MC2, Bridge-B2, Bridge E, G0.11-0.11 & 6645 &   2006-09-25 &  Sgr A$^{*}$ &  17 45 40.00  -29 00 28.00 & 5\\
MC2, Bridge-B2, Bridge E, G0.11-0.11 & 6646 &   2006-10-29 &  Sgr A$^{*}$ &  17 45 40.00  -29 00 28.00 & 5\\
\hline
MC2, Bridge-B2, Bridge E, G0.11-0.11 & 7554 &   2007-02-11 &  Sgr A$^{*}$ &  17 45 40.00  -29 00 28.00 & 5\\
MC2, Bridge-B2, Bridge E, G0.11-0.11 & 7555 &   2007-03-25 &  Sgr A$^{*}$ &  17 45 40.00  -29 00 28.00 & 5\\
MC2, Bridge-B2, Bridge E, G0.11-0.11 & 7556 &   2007-05-17 &  Sgr A$^{*}$ &  17 45 40.00  -29 00 28.00 & 5\\
MC2, Bridge-B2, Bridge E, G0.11-0.11 & 7557 &   2007-07-20 &  Sgr A$^{*}$ &  17 45 40.00  -29 00 28.00 & 5\\
MC2, Bridge-B2, Bridge E, G0.11-0.11 & 7558 &   2007-09-02 &  Sgr A$^{*}$ &  17 45 40.00  -29 00 28.00 & 5\\
MC2, Bridge-B2, Bridge E, G0.11-0.11 & 7559 &   2007-10-26 &  Sgr A$^{*}$ &  17 45 40.00  -29 00 28.00 & 5\\
\hline
MC2, Bridge-B2, Bridge E, G0.11-0.11 & 9169 &   2008-05-05 &  Sgr A$^{*}$ &  17 45 40.00  -29 00 28.10 & 28\\
MC2, Bridge-B2, Bridge E, G0.11-0.11 & 9170 &   2008-05-06 &  Sgr A$^{*}$ &  17 45 40.00  -29 00 28.10 & 27\\
MC2, Bridge-B2, Bridge E, G0.11-0.11 & 9171 &   2008-05-10 &  Sgr A$^{*}$ &  17 45 40.00  -29 00 28.10 & 28\\
MC2, Bridge-B2, Bridge E, G0.11-0.11 & 9172 &   2008-05-11 &  Sgr A$^{*}$ &  17 45 40.00  -29 00 28.10 & 27\\
MC2, Bridge-B2, Bridge E, G0.11-0.11 & 9174 &   2008-07-25 &  Sgr A$^{*}$ &  17 45 40.00  -29 00 28.10 & 29\\
MC2, Bridge-B2, Bridge E, G0.11-0.11 & 9173 &   2008-07-26 &  Sgr A$^{*}$ &  17 45 40.00  -29 00 28.10 & 28\\
\hline
MC2, Bridge-B2, Bridge E, G0.11-0.11 & 10556 & 2009-05-18 &  Sgr A$^{*}$ &  17 45 40.00  -29 00 28.10 & 113\\
\hline
MC2, Bridge-B2, Bridge E, G0.11-0.11 & 11843 & 2010-05-13 &  Sgr A$^{*}$ &  17 45 40.00  -29 00 28.00 & 79\\
\hline
MC2, Bridge-B2, Bridge E, G0.11-0.11 & 13016 & 2011-03-29 &  Sgr A$^{*}$ &  17 45 40.00  -29 00 28.10 & 18\\
MC2, Bridge-B2, Bridge E, G0.11-0.11 & 13017 & 2011-03-31 &  Sgr A$^{*}$ &  17 45 40.00  -29 00 28.10 & 18\\
MC2, Bridge-B2, Bridge E, G0.11-0.11 & 13508 & 2011-07-19 &  Sgr A complex &  17 45 59.70  -28 58 15.90 & 33\\
MC2, Bridge-B2, Bridge E, G0.11-0.11 & 12949 & 2011-07-21 &  Sgr A complex &  17 45 59.70  -28 58 15.90 & 58\\
MC2, Bridge-B2, Bridge E, G0.11-0.11 & 13438 & 2011-07-29 &  Sgr A complex &  17 45 59.70  -28 58 15.90 & 66\\
\hline
MC2, Bridge-B2, Bridge E, G0.11-0.11 & 14941 & 2013-04-06 &  Sgr A$^{*}$ &  17 45 40.00  -29 00 28.10 & 20\\
MC2, Bridge-B2, Bridge E, G0.11-0.11 & 14942 & 2013-04-14 &  Sgr A$^{*}$ &  17 45 40.00  -29 00 28.10 & 20\\
\hline
MC2, Bridge-B2, Bridge E, G0.11-0.11 & 17236 & 2015-04-25 &  Sgr A complex 1 &  17 46 15.50  -28 55 00.70 & 79\\
MC2, Bridge-B2, Bridge E, G0.11-0.11 & 17239 & 2015-08-19 &  Sgr A complex 2 &  17 46 07.00  -28 53 09.50 & 79\\
\hline
MC2, Bridge-B2, Bridge E, G0.11-0.11 & 17237 & 2016-05-18 &  Sgr A complex 1 &  17 46 15.50  -28 55 00.70 & 21\\
MC2, Bridge-B2, Bridge E, G0.11-0.11 & 18852 & 2016-05-18 &  Sgr A complex 1 &  17 46 14.10  -28 54 52.50 & 52\\
MC2, Bridge-B2, Bridge E, G0.11-0.11 & 17240 & 2016-05-18 &  Sgr A complex 2 &  17 46 09.60  -28 53 43.80 & 75\\
\hline
MC2, Bridge-B2, Bridge E, G0.11-0.11 & 17238 & 2017-07-17 &  Sgr A complex 1 &  17 46.14.10  -28 54 52.50 & 65\\
MC2, Bridge-B2, Bridge E, G0.11-0.11 & 20118 & 2017-07-23 &  Sgr A complex 1 &  17 46 14.10  -28 54 52.50 & 14\\
MC2, Bridge-B2, Bridge E, G0.11-0.11 & 17241 & 2017-10-02 &  Sgr A complex 2 &  17 46.07.00  -28 53 09.50 & 25\\
MC2, Bridge-B2, Bridge E, G0.11-0.11 & 20807 & 2017-10-05 &  Sgr A complex 1 &  17 46.07.00  -28 53 09.50 & 28\\
MC2, Bridge-B2, Bridge E, G0.11-0.11 & \textbf{20808}\tablefootmark{*} & 2017-10-02 &  Sgr A complex 2 &  17 46.07.00  -28 53 09.50 & 27\\
\hline
\hline
Sgr B2 & \textbf{11795} \tablefootmark{*}& 2010-07-20 &  Sgr B2  &  17 46 06.70  -28 26 47.29 & 99\\
\hline
\hline
Sgr C1, Sgr C2 & \textbf{16643}\tablefootmark{*} & 2014-08-03 & Sgr C  &  17 44 23.80  -29 23 58.90 & 36\\
Sgr C3 &  \textbf{7040}\tablefootmark{*} & 2007-04-25 &  Deep GCS 8 &  17 45 25.10  -29 23 33.60 & 37\\
\hline
\end{tabular}
\tablefoot{
\tablefoottext{*}{Observations that we use for the spectral analysis.}}
\end{table*}

\end{appendix}

\end{document}